\documentclass[sigconf]{acmart}
\usepackage{latexsym}

\usepackage{booktabs}
\usepackage{array}
\usepackage{multirow}
\usepackage{graphicx}
\usepackage{color}
\usepackage{blindtext}
\usepackage{caption}
\usepackage{subcaption}
\usepackage{enumitem}
\usepackage{multirow}
\usepackage{tikz}
\usepackage{amsmath,bm,amssymb}
\usepackage{comment}
\usepackage{tabulary}
\usepackage{makecell}
\usepackage{tcolorbox}
\usepackage{cleveref}
\usepackage{balance}
\setcopyright{rightsretained}

\copyrightyear{2018}
\acmYear{2018} 
\setcopyright{iw3c2w3}
\acmConference[WWW 2018]{The 2018 Web Conference}{April 23--27, 2018}{Lyon, France}
\acmBooktitle{WWW 2018: The 2018 Web Conference, April 23--27, 2018, Lyon, France}
\acmPrice{}
\acmDOI{10.1145/3178876.3186142}
\acmISBN{978-1-4503-5639-8/18/04}


\newif{\ifhidecomments}
\hidecommentsfalse

\ifhidecomments
    \newcommand{\chenhao}[1]{}
    \newcommand{\hao}[1]{}
    \newcommand{\nascomment}[1]{}
\else
    \newcommand{\chenhao}[1]{\textcolor{blue}{[#1 --\textsc{ct}]}}
    \newcommand{\hao}[1]{\textcolor{red}{[#1 --\textsc{hp}]}}
    \newcommand{\nascomment}[1]{\textcolor{brown}{[#1 --\textsc{nas}]}}
\fi

\newcommand{\addFigure}[2]{\includegraphics[width=#1]{plots/#2}}
\newcommand{\para}[1]{\noindent{\bf #1}}
\newcommand{\figref}[1]{Figure~\ref{#1}}
\newcommand{\secref}[1]{\S\ref{#1}}
\newcommand{\tableref}[1]{Table~\ref{#1}}
\newcommand{\positive}{highlighted\xspace}
\newcommand{\negative}{not-highlighted\xspace}
\newcommand{\Positive}{Highlighted\xspace}

\newcommand{\highlight}{highlight\xspace}
\newcommand{\highlighted}{highlighted\xspace}
\newcommand{\highlights}{highlights\xspace}

\newcommand{\notbow}{{``all -- BOW''}\xspace}

\definecolor{positivecolor}{RGB}{1, 56, 106}
\definecolor{negativecolor}{RGB}{55, 120, 191}
\newcommand{\positivesent}[1]{{\bf \textcolor{positivecolor}{#1}}\xspace}
\newcommand{\negativesent}[1]{{\bf \textcolor{negativecolor}{#1}}\xspace}
\newcommand{\msmatrix}{\mathbf{D}}

\author{Chenhao Tan}
\affiliation{%
   \department{Department of Computer Science}
  \institution{University of Colorado Boulder}
  \city{Boulder} 
  \state{CO} 
  \postcode{80309}
  \country{USA}
}
\email{chenhao@chenhaot.com}
\author{Hao Peng}
\affiliation{%
   \department{Paul G. Allen School of CS\&E}
  \institution{University of Washington}
  \city{Seattle} 
  \state{WA} 
  \postcode{98195}
  \country{USA}
}
\email{hapeng@cs.washington.edu}
\author{Noah A. Smith}
\affiliation{%
   \department{Paul G. Allen School of CS\&E}
  \institution{University of Washington}
  \city{Seattle} 
  \state{WA} 
  \postcode{98195}
  \country{USA}
}
\email{nasmith@cs.washington.edu}

\title{``You are no Jack Kennedy'': \\ 
On Media Selection of
Highlights from
Presidential Debates
}


\renewcommand{\shorttitle}{``You are no Jack Kennedy'': 
On Media Selection of
Highlights from
Presidential Debates}
\renewcommand{\shortauthors}{Tan, Peng, and Smith}

\begin{document}

\begin{abstract}

Political speeches and debates play an important role in shaping the images of politicians, and the public often relies on media outlets to select bits of political communication from a large pool of utterances.
It is an important research question to understand what factors impact this selection process.

To quantitatively explore the selection process,
we build a three-decade dataset of presidential debate transcripts and post-debate coverage.
We first examine the effect of wording and propose a binary
classification framework that controls for both the speaker and the debate situation.
We find that crowdworkers can only achieve an accuracy of 60\% in this task,
indicating that media choices are not entirely obvious. 
Our classifiers 
outperform crowdworkers on average, mainly 
in primary debates.
We also compare important factors from crowdworkers' free-form explanations with those from data-driven methods and find interesting differences.
Few crowdworkers mentioned that ``context matters'',
whereas our data show 
that well-quoted sentences 
are more distinct from the previous utterance by the same speaker
than less-quoted sentences.
Finally, we examine the aggregate effect of media preferences towards different wordings to understand the extent of fragmentation among media outlets. 
By analyzing a bipartite graph built from quoting behavior in our data,
we observe a decreasing trend in bipartisan coverage.
\end{abstract}
    \maketitle

\section{Introduction}
\label{sec:intro}

Televised public debates have
become a focal point of election campaigns
\cite{mckinney2004political}. 
A famous example is from the 1988 U.S. vice presidential debate.
After Dan Quayle compared himself to John F. Kennedy,
Lloyd Bentsen 
dismissively replied, ``you are no Jack Kennedy''.
This moment received wide post-debate coverage, and even pervades later debates and popular parodies.\footnote{\url{https://en.wikipedia.org/wiki/Senator,_you're_no_Jack_Kennedy\#Legacy}.}
We refer to  moments that are frequently quoted by media outlets
as {\em \highlights}.  
Media-selected \highlights in post-debate coverage shape how the public interprets election debates, because these \highlights may be the only debate content consumed by many voters \cite{Fridkin01012008,Hillygus:AmericanJournalOfPoliticalScience:2003,hwang2007applying}.
\begin{figure}[t]
\centering
\vspace{0.4cm}
{
\begin{tabular}{p{0.5\textwidth}}
\begin{tcolorbox}[width=0.45\textwidth,
                  boxsep=0pt,
                  left=5pt,
                  right=5pt,
                  top=3pt,
                  bottom=3pt,
                  arc=8pt,
                  boxrule=1.5pt,
                  colback=gray!20!white
                  ]
{\bf SANDERS}: \positivesent{Do I consider myself part of the casino capitalist process by which so few have so much and so many have so little by which Wall Street's greed and recklessness wrecked this economy?} [...]\\
{\bf CLINTON}: 
[...] I think what Senator Sanders is saying certainly makes sense in the terms of the inequality that we have. [...]\\
{\bf SANDERS}: [...]
\negativesent{So what we need to do is support small and medium-sized businesses, the backbone of our economy, but we have to make sure that every family in this country gets a fair shake} [...]
\end{tcolorbox}
\end{tabular}
}
\vspace{-0.4cm}
\caption{
Between the two bold sentences from Bernie Sanders from neighboring turns in the 2016 Democratic primary debates, the first sentence was quoted 23 times 
in newspapers within a week after the debate, while the second one was
not quoted at all in our data. 
Yet in our experiments,
3 out of 5 humans 
thought the second one was quoted more.
}
\label{tab:example}
\end{figure}

However, most \highlights that media select are not as exceptional as ``you are no Jack Kennedy'',
and it remains unclear what factors determine media selection.
Consider the example in \figref{tab:example}.
It was not obvious to 
participants in our experiments
which of the two passages from Sanders was a \highlight.
Even knowing that the first one was \highlighted,
we can propose multiple plausible explanations for this choice of the media.
It could 
be the catchiness of ``casino capitalist'', or the parallel structure of ``so few have so much'' and ``so many have so little''.
It may also relate to the conversational dynamics (e.g., 
Clinton's agreement about inequality) or non-textual factors such as the media's political leanings.

Some
qualitative
studies have 
investigated the effect of language-related factors in how the media select highlights
\cite{atkinson1984our,clayman1995defining,hallin1992sound,gidengil2003talking}.
For instance, to explain the popularity of ``you are no Jack Kennedy'', \citet{clayman1995defining} suggests three important factors: 
1) {\em narrative relevance} (how well a moment fits in a news story); 2) {\em conspicuousness} (how much a moment stands out in a debate); 3) {\em extractability} (how self-contained a moment is).
However, it is nontrivial to computationally characterize these qualitative factors, and their predictive power remains unknown.
What is also missing in the existing literature is an understanding of how consumers of news coverage, i.e., the public, interpret media outlets' selection of highlights. 

Moreover, media selection of highlights holds promise for understanding media bias and polarization.
Existing studies have shown that 
non-textual factors such as
the media's preferences and biases can affect the selection of \highlights
\cite{groseclose2005measure,lin2011more,Baum:PoliticalCommunication:2008,Niculae:2015:QSP:2736277.2741688}.
In particular, \citet{Niculae:2015:QSP:2736277.2741688} demonstrate implicit structure in the media (e.g., international vs. domestic) by analyzing the patterns in quotes of President Barack Obama.
Since televised presidential debates have been happening for decades, analysis of debate coverage can shed light on the evolution of media preferences over time.

To quantitatively investigate these questions,
we collect American presidential debate transcripts,
including both general debates (the debates between general election
candidates after primary elections) and primary debates (the
earlier, within-party debates in primary elections), and post-debate coverage in newspapers
 (details in \secref{sec:data}). 
Our dataset spans more than three decades.

\para{The present work: the effect of wording on media choices (\secref{sec:textual}).}
The first thrust of this paper investigates the effect of wording on media choices and examines whether the public understands these choices.
To do that, we propose a binary classification framework, 
where a well-quoted sentence (highlight) is paired with a 
sentence that is not a highlight, 
controlling for the speaker and the debate situation.
The task is to identify which one was quoted more.
Using this classification framework, we investigate how well humans and machine-learned classifiers can predict media choices and what the distinguishing textual factors are.

We find that media choices in the selection of highlights are not entirely obvious to humans.
As a proxy for the general public, we request 
Mechanical Turk 
workers
to perform the classification task and explain what factors they use in making predictions.
Although they are able to identify some textual signals
and outperform random chance (50\%),
they only achieve an average accuracy of 60\%.
Meanwhile, there seem to exist more signals in the wording that are 
not salient to untrained humans.
With carefully-designed features 
that
build on past qualitative studies
\cite{atkinson1984our,clayman1995defining,hallin1992sound,gidengil2003talking},
our classifiers achieve an accuracy of 66\%.
This result indicates that textual factors can predict
media choices to a greater extent than average human performance suggests.
In fact, the main performance gap 
comes from primary debates. 
One possible explanation for the gap in human performance between general debates and primary debates 
is the amount of past exposure:  
primary debates receive less media coverage than general debates and humans
may have weaker memories of primary debate highlights.
We also observe interesting similarities and differences when comparing 
distinguishing factors that humans mentioned 
with 
those identified by data-driven methods.
For instance,
negativity is
considered important in both approaches.
However, the two approaches view conversational context differently.
Only 3\% of human responses mention that context matters,
while 
our models suggest that it is a significant factor: 
\highlights tend to be more different from the speaker's previous utterance, 
and are more likely to be picked up in later utterances.

\para{The present work: quoting patterns over time (\secref{sec:media_preferences}).}
The second thrust of this paper examines the media's own preferences and biases in selecting \highlights over time.
Instead of viewing all media outlets as a uniform body, we take advantage of the longitudinal nature of our data and study whether the news media have 
become more fragmented over time.
Using a bag-of-sentences approach, 
we construct a bipartite graph over media outlets and the sentences they quoted.
Consistent with existing studies on polarization \cite{Baum:PoliticalCommunication:2008}, we observe a decreasing trend in bipartisan coverage in general elections, where a clear two-party structure exists.
When we investigate the similarity between media outlets without partisan assumptions,
we find an increasing trend in the tightness of local clustering, but do not observe that media outlets 
are becoming less similar to each other over time on
average. 

\begin{figure*}[t]
    \centering
    \begin{subfigure}[t]{0.23\textwidth}
        \addFigure{\textwidth}{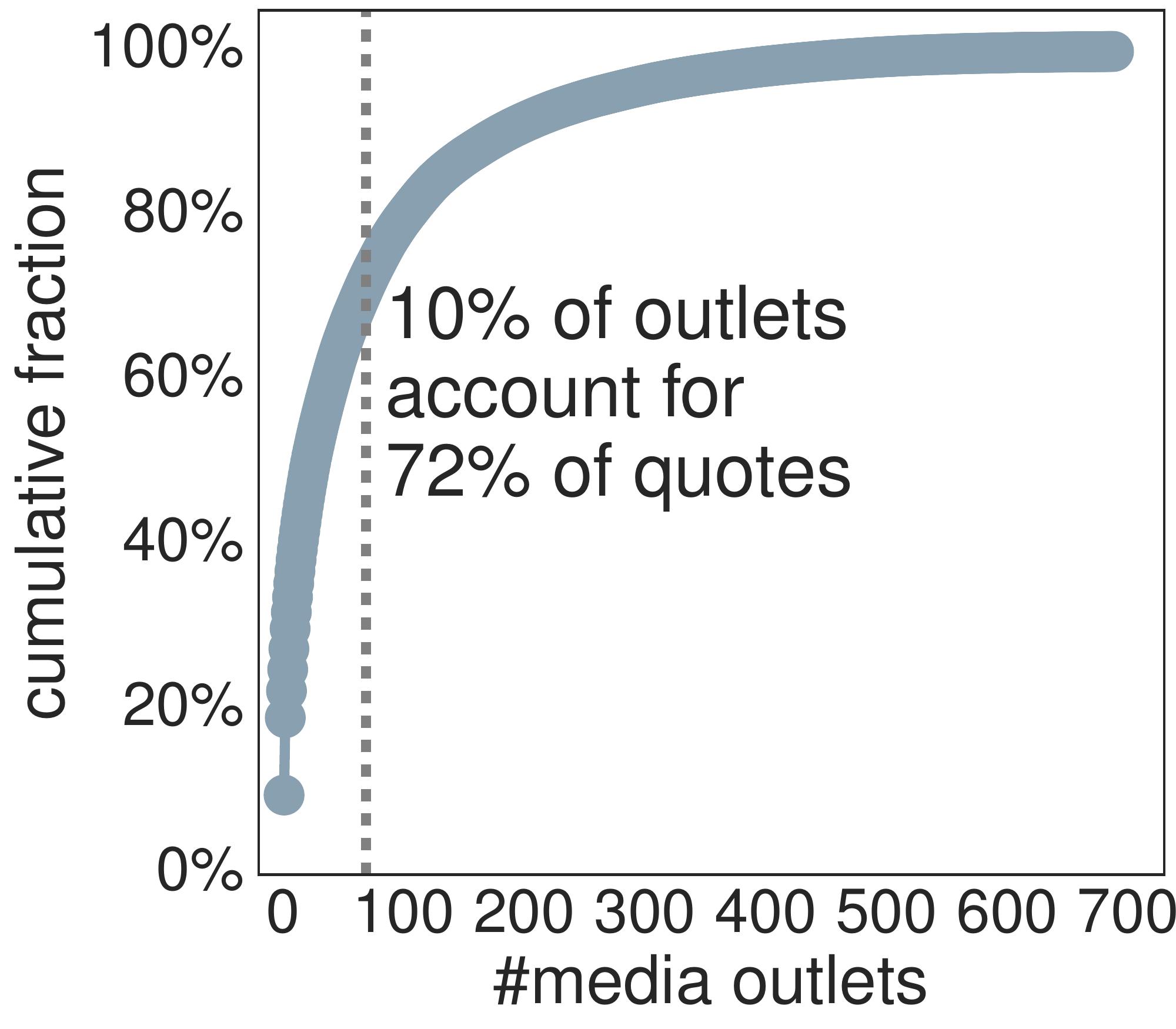}
        \caption{Cumulative fraction of quotes by media outlet.}
        \label{fig:news_dist}
    \end{subfigure}
    \hfill
    \begin{subfigure}[t]{0.23\textwidth}
        \addFigure{\textwidth}{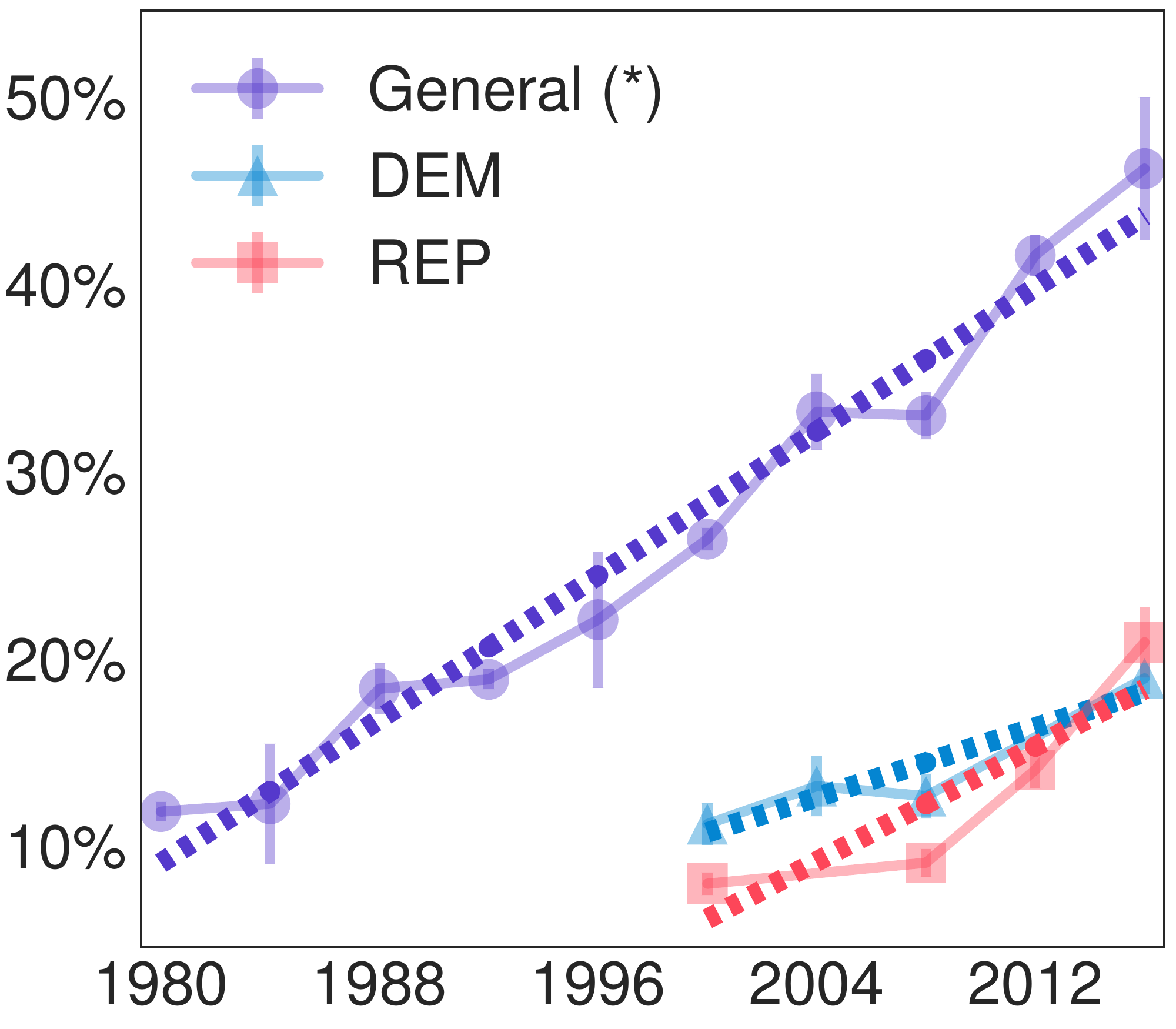}
        \caption{Fraction of debate sentences being (partially) quoted.}
        \label{fig:speech_quoted}
    \end{subfigure}
    \hfill
    \begin{subfigure}[t]{0.23\textwidth}
        \addFigure{\textwidth}{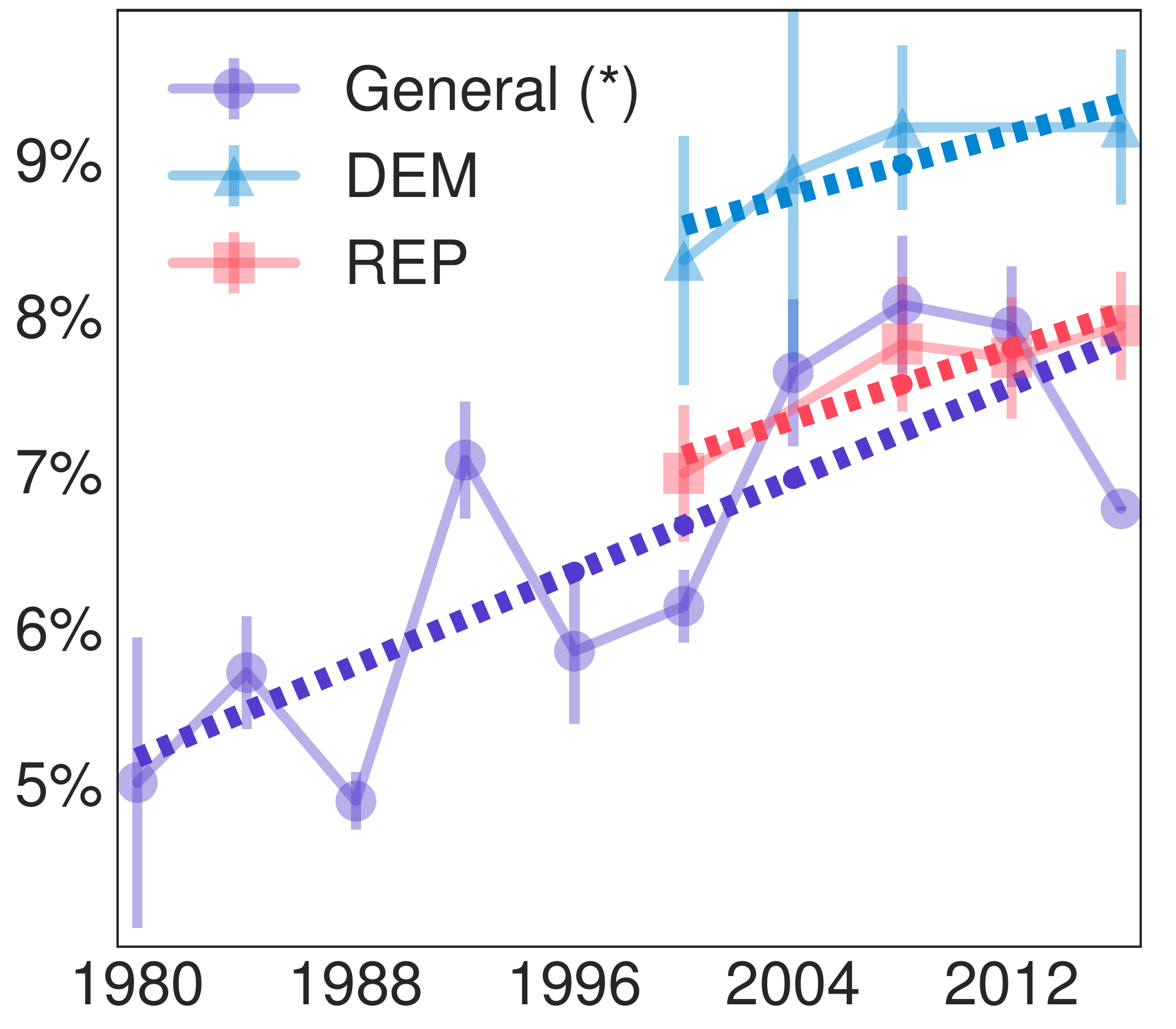}
        \caption{Fraction of texts that are quotes in news articles.}
        \label{fig:article_quotes}
    \end{subfigure}
    \hfill
    \begin{subfigure}[t]{0.23\textwidth}
        \addFigure{\textwidth}{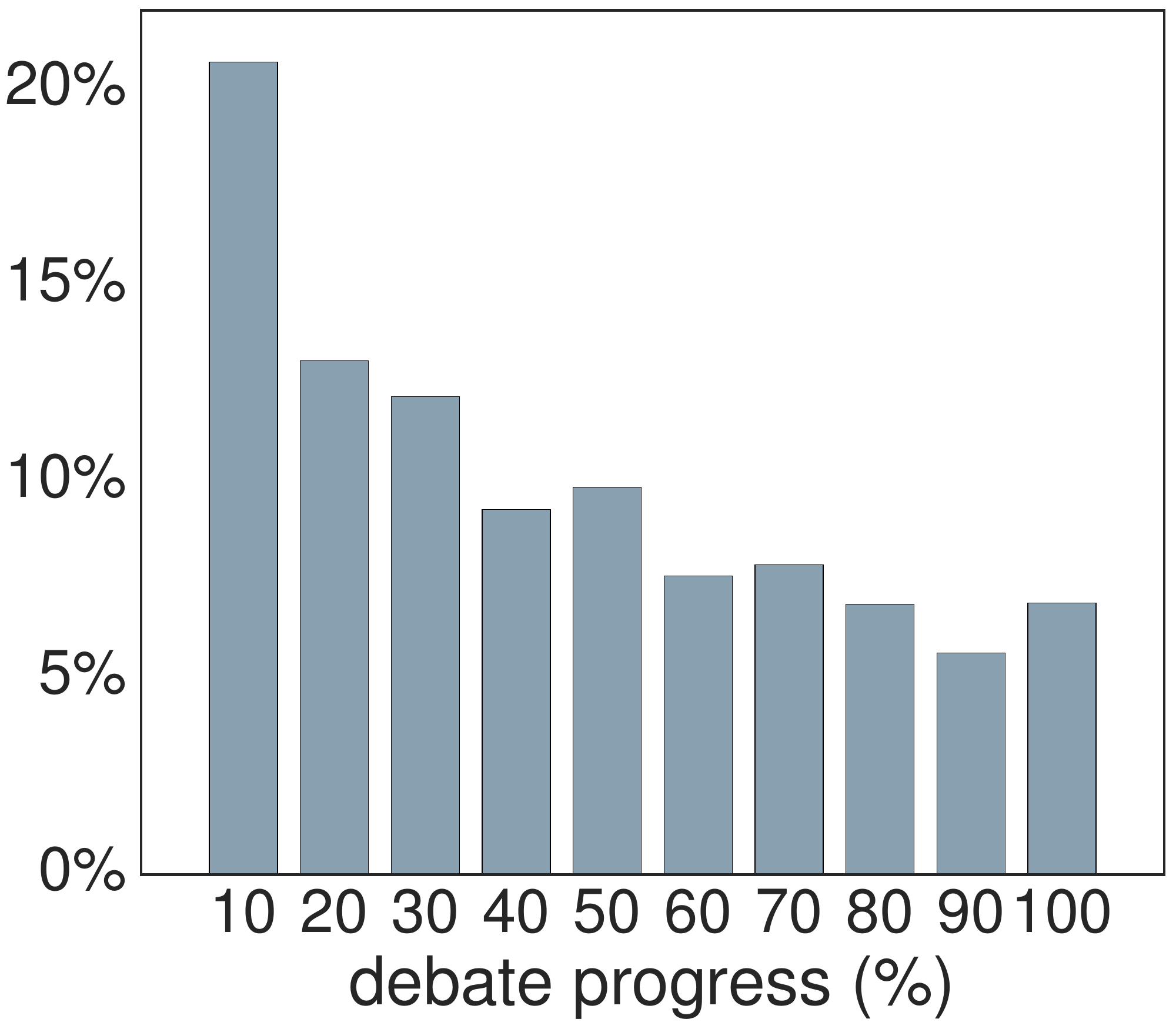}
        \caption{Fraction of quotes from each decile of turns.}
        \label{fig:quote_dist}
    \end{subfigure}
    \caption{ 
    In \figref{fig:news_dist}, media outlets are sorted by total number of quotes, and there is a heavy tail.
    \figref{fig:speech_quoted} 
    shows that 
    an increasing fraction of sentences in the debates are (partially) quoted by the media over time.
    \figref{fig:article_quotes} indicates that an increasing fraction of texts in news articles are direct quotes from the debates.
    \figref{fig:quote_dist} shows that more quotes are from the beginning of a debate.    
    Throughout the paper, error bars represent standard error, dashed lines show the best linear fit and * in a legend indicates that 
    the linear coefficient is statistically significantly different from 0 with $p<0.05$.
    }
    \label{fig:fraction_cmp}
\end{figure*}

\section{An Overview of the Dataset}
\label{sec:data}

Our dataset consists of two parts: debate transcripts and 
post-debate news coverage.
We extract transcripts from general debates since 1960
and primary debates since the 2000 presidential election from the \textit{American Presidency Project}.\footnote{\url{http://www.presidency.ucsb.edu/debates.php}.
In fact, presidential debates are a relatively recent phenomenon, despite their prominence nowadays. After the first general debate between John F. Kennedy and Richard Nixon in 1960, there were no general debates until 1976. For more historical details, refer to \url{http://www.cnn.com/2012/09/30/opinion/greene-debates/}.}
Throughout this work, 
we define a {{\em turn}} as an uninterrupted utterance by a single speaker.

In order to collect 
post-debate news coverage, we use LexisNexis Academic\footnote{\url{http://www.lexisnexis.com/hottopics/lnacademic/}.} to search all newspapers within seven days after each debate.
As LexisNexis indexes newspapers since 1980, we study media highlights of presidential debates from 1980 to 2016.
To achieve high recall, we use debate type (``presidential'', ``vice'', ``democratic'' and ``republican'') and the word ``debate'' as the query.
Although newspapers are a subset of news coverage, 
they comprise a long-standing and often-studied segment of the media
and are highly amenable to replication studies.
We leave exploration of additional media sources to future work.

Inspired by existing studies \cite{leskovec2009meme,Niculae:2015:QSP:2736277.2741688,simmons2011memes,tan2016lost}, 
we define ``highlights'' 
based on quotes in news articles that directly come from the debates. 
In this work, we differentiate quotes from quotations.
We refer to any texts in news articles that are enclosed in quotation marks  as {\em quotations}, and {\em quotes} are a subset of quotations that can be matched to a turn at a debate.
We determine whether a quotation matches a turn in the presidential debate based on word overlap and fuzzy matching.
The extraction process
produces pairs of quotes and quoted sentences from the corresponding debate.
We will present the formal definition of \highlights in \secref{sec:quotability}.
Our dataset and supplementary material are available at \url{https://chenhaot.com/papers/debate-quotes.html}.

\tableref{tab:basic_stats} shows overall
statistics of our dataset.
We next discuss basic properties of our dataset. In particular, we observe an increasing trend of news media quoting presidential debate moments.

\para{A diverse set of newspapers (\figref{fig:news_dist}).}
It is important to point out that there 
are more newspapers over time in our dataset, partly because more media outlets began to quote presidential debates and partly because LexisNexis gradually improves 
their collection
of newspapers.
Only four newspapers 
quoted general debates in 1980; 334 newspapers 
did in 2016.
There are around 700 unique newspapers in total and the top 10\% of the newspapers in quoting debates account for 72\% of all the quotes.
The \emph{New York Times} and \emph{Washington Post} are consistently the top newspapers with the most quotes since 1980.
We also have small newspapers (e.g., \emph{Rhode Island Lawyers Weekly}) and international newspapers (e.g., \emph{The Guardian}).

\begin{table}
\centering
\begin{tabular}{lrp{0.4in}p{0.4in}p{0.4in}}
\toprule
debate type & \#debates & avg. \#sent's & avg. \#tokens & avg. \#quotes\\
\midrule
general & 26 & 1064.9 & 16278.0 & 944.2\\
vice & 9 & 1018.2 & 15974.0 & 618.4\\
Democratic & 38 & 1070.6 & 16028.3 & 330.7\\
Republican & 59 & 1270.8 & 17781.1 & 369.1\\
\bottomrule
\end{tabular}

\caption{Dataset statistics. The last three columns show the average number of sentences, the average number of tokens, and the average number of quotes per debate, respectively.}
\label{tab:basic_stats}
\end{table}

\para{An increasing trend of quoting (\figref{fig:speech_quoted} and \figref{fig:article_quotes}).}
Since there are more media outlets over time,
it is expected that an increasing fraction of sentences in the debates are quoted in the news media.
In comparison, general debates are quoted much more than primary debates.
But it is unexpected that the fraction of texts that are direct quotes
in news articles is also growing over time, as we observe.
This suggests that directly quoting the candidates is an increasingly common way to cover debates.

\para{More quotes are from the beginning of a debate (\figref{fig:quote_dist}).}
Later turns in a debate are less likely to be quoted in the media.
This decreasing likelihood 
is robust across different types of debates and echoes findings on movie quotes \cite{Danescu-Niculescu-Mizil+Cheng+Kleinberg+Lee:12}.

\section{The Effect of Wording on Media Choices}
\label{sec:textual}

We 
study how 
textual factors
associate with
media selection of highlights from presidential debates for three reasons.
First, 
media-selected highlights are the only debate content consumed by voters  
who do not watch the debate and hence rely on post-debate coverage.
How well the public understands media selection of highlights is worth studying because the public uses this understanding to interpret news coverage.
Second, 
debating candidates cannot control their popularity 
or the news media's political preferences, 
but they can always choose the wording when they seek to deliver a message.
Understanding the effects of wording can thus inform political communication.
Third, it is valuable to know the extent to which we are able to
predict media choices using {\em only} textual information---{\em although
of course textual information alone may not fully predict media choices}.

To study the effect of wording on media choices, we propose an experimental framework that controls for the speaker and the debate situation
and formulate a binary classification task (\secref{sec:quotability}).
We then study the public understanding of media choices by evaluating human performance on this task and analyzing free-form explanations in human surveys (\secref{sec:human_eval}). 
We further build on existing theories and develop quantitative features for data-driven classifiers (\secref{sec:machine_features}), and examine their prediction performance in \secref{sec:performance}.

\subsection{Experimental Framework}
\label{sec:quotability}

To investigate 
how textual factors associate with
media selection of \highlights from presidential debates, 
we need to control for other confounding factors such as who the speaker is and what state the debate is in. 
Inspired by ``natural experiments'' and previous studies about the effect of wording on message sharing, memorability, and persuasion \cite{Danescu-Niculescu-Mizil+Cheng+Kleinberg+Lee:12,Dinardo:Microeconometrics:2010,Tan:ProceedingsOfAcl:2014,Tan:2016:WAI:2872427.2883081},
we propose a classification task that asks humans and machines to decide which sentence was quoted more in the news media between two ``similar'' sentences.

\para{Binary classification framework.} 
To formally define \highlights, 
we use 
a sentence as the basic unit of analysis.
In our natural experiment framework, we find a matching ``negative'' sentence for each 
media-selected \highlight
and evaluate whether humans or machines can tell the \positive one from the \negative one in a pair.
Because 
debating candidates 
have varying popularity and the debate progresses 
with different levels of importance (see \figref{fig:quote_dist}),
we match each 
\highlight with a \negative sentence of {similar length}  within 
three turns by the {same speaker}.\footnote{There exist alternate ways to account for topic shift, e.g., \cite{Nguyen:2012:SHN:2390524.2390536}.}
We consider a sentence {\em \positive} if it was among the most quoted
$t\%$ sentences from the corresponding debate.
We opted for this instead of absolute-count thresholding because
the number of quotes increases over time
(see \figref{fig:speech_quoted}).
We experiment with $t=1,2,\dots, 10$.
The results are robust across choices of $t$, and we thus report only the results for $t=10$ (except for overall accuracy). %

Following the above definition, we extract $\sim$14K pairs of sentences from all the debates.
For a pair of sentences, we randomize the order and predict whether the first one was \positive.
A random guess gives an accuracy of 50\%.
We randomly select 80\% of our data for training and hold out the other 20\% for testing.
To build machine learning classifiers in \secref{sec:performance},
we construct a vector representation of each pair by extracting features from each sentence and take the difference between them.
We use logistic regression with $\ell_2$-regularization.
This approach is equivalent to a linear framework for the ranking task within a pair \cite{Joachims:2002:OSE:775047.775067}.
We grid search 
the best $\ell_2$ coefficient based on five-fold cross-validated accuracy on the training set over
$\bigl\{2^x\bigr\}$, where $x$ ranges over 20 values evenly spaced between --8 and 1.

\begin{table}[t]
    \centering
\begin{tabular}{p{0.32\textwidth}r}
\toprule
category &  \% humans \\
\midrule
circular (sound bite, newsworthy) & 30.0 \\
\midrule
provocative, sensational &  25.5 \\
surprising, funny &  17.0\\
issues, informative & 16.0 \\
controversial &  15.0 \\
memory, past exposure & 12.0 \\
\bottomrule
\end{tabular}

    \caption{Top factors in human surveys and the percentage of humans that mentioned them.}
    \label{tab:human_factors}
\end{table}

\subsection{Human Interpretation of Media Choices}
\label{sec:human_eval}

Using the above classification framework, we first 
examine the public understanding of how the news media select \highlights.
We recruit 200 U.S.-based 
Mechanical Turk workers as a sample of 
untrained humans (the public) to perform the prediction task on randomly sampled pairs from the held-out set.
In addition, we ask our participants to explain the important factors that they use to make predictions in free-form responses.

Specifically, we request each participant to label 25 pairs and finish
an exit survey to explain what factors they used to make predictions
as well as their experience of watching debates and their political ideology.
For a pair, we show the \positive sentence, the \negative sentence, and a few surrounding sentences\footnote{For both sentences in a pair, we include 
up to 
3 sentences before it and 3 sentences after it to provide some context for the participants to understand the current state of the debate.} in the order 
they occurred in the debate and ask the participants to guess which one was quoted more in the news media.
To make sure that they understand the task, we prepare three training
pairs and require a comprehension quiz before they start. 
We also provide a bonus for each correct guess 
to incentivize participants to try their best.
Further details of the human experiments are in the appendix.
\para{Media choices are not obvious to the public.}
The average human accuracy is 60\% and Fleiss' $\kappa$ between human labels is 0.2, indicating slight agreement.\footnote{Surprisingly, 
there is no clear relationship between an individual's
prediction performance and their self-reported level of experience or political
ideology.}
These observations suggest that media choices are not obvious to humans, at least based on textual content.
One plausible explanation is that the textual information is insufficient to explain media choices: media choices are influenced by external factors such as statements made outside presidential debates and public opinion shifts.
However, as we will show later, there seem to
exist signals in the wording that are not salient to untrained
humans. 
Another more pessimistic explanation is that humans have a limited understanding of how the news media select highlights.

\para{Important factors in human surveys.} To examine the important factors from the perspective of humans, 
we categorize free-form explanations in human surveys and present the top factors in \tableref{tab:human_factors}.
The most common factors cited are circular;
i.e., 30\% of the participants mentioned that they made decisions based
on which one is newsworthy or which one makes a good sound bite.
This suggests that it is nontrivial for humans to reason about media selection of highlights.
Among the next five most frequently mentioned categories, participants mentioned {\em sensational} (emotional, negative, shocking, etc.) and {\em surprising or funny}.
Most of these factors are difficult to operationalize computationally.
Interestingly, {\em memory or past exposure} was explicitly mentioned by 12\% of the participants, 
indicating that humans may predict media choices even less accurately without unavoidable media exposure.

%
%
%
%
%
%
%
%
%
%
%

These top factors do not directly align with 
existing qualitative studies.
For instance, \citet{clayman1995defining}, the most relevant work, points out three important factors: 
1) {\em narrative relevance} (how well a moment fits in a news story); 2) {\em conspicuousness} (how much a moment stands out in a debate); 3) {\em extractability} (how self-contained a moment is).
It is unclear how to map the factors that our participants mentioned to these three.

Notably, only 3\% of the participants mentioned that context in which a sentence occurred matters,
while an equal number of people 
brought up
appealing to liberal voters as a criterion (none discussed the other direction).
Extractability in \citet{clayman1995defining}, or ``can be taken out of context'' was viewed important by 4\% of the participants.
However, ``potential to be twisted'', a slightly different but more malicious interpretation, was explicitly mentioned by 6.5\% of the participants.
These observations indicate a negative attitude toward the news media,
or at least some skepticism about their role in American
politics.
%
%
%

%
%

%

%

%
%
\begin{table*}[t]
	\begin{center}
		\begin{tabulary}{\textwidth}{@{}p{0.1\textwidth}@{\hspace{2pt}} @{\hspace{2pt}} p{0.72\textwidth}  @{\hspace{2pt}}p{0.12\textwidth} @{\hspace{2pt}}r@{}} 

			\toprule
			Feature set & \multicolumn{1}{c}{Related theories/intuitions and brief description} & \multicolumn{2}{c}{Significance} \\
			\midrule
			\multirow{2}{\hsize}{Informative-ness}
			& \multirow{2}{\hsize}{We use length as a proxy of informativeness. 
			Longer sentences are more likely to be
                          \highlighted despite our control on length as discussed in \secref{sec:quotability}.
    		This echoes findings in \citet{Tan:ProceedingsOfAcl:2014,Tan:2016:WAI:2872427.2883081}.}
			& \multirow{2}{*}{length} & \multirow{2}{*}{$\uparrow\uparrow\uparrow\uparrow$}\\
			& & & \\
			\noalign{\smallskip}
			\hline

			\multirow{3}{*}{Emotions}
			& \multirow{3}{\hsize}{We consider positive and negative words in \citet{pennebaker2001linguistic}.
    		\Positive sentences use significantly more negative words, while there is no difference in positive words. This is consistent with negativity bias \cite{rozin2001negativity} and the negativity found in the news media \cite{geer2012news}.}
			& posemo & \\ 
			& &	negemo & $\uparrow\uparrow\uparrow\uparrow$ \\
			& & & \\
			\noalign{\smallskip}
			\hline

			\multirow{2}{*}{Contrast}
			& \multirow{2}{\hsize}{%
			We use negations and negative conjunctions (e.g., \emph{not, but, although}) to capture contrast.
			Our result echoes \citet{atkinson1984our}, which demonstrates the importance of contrast.}
			& negation & $\uparrow\uparrow\uparrow\uparrow$\\
			& & negative conj. & $\uparrow\uparrow\uparrow\uparrow$\\
			\noalign{\smallskip}
			\hline

			\multirow{3}{\hsize}{Personal pronouns}
			& \multirow{3}{\hsize}{In general, \positive sentences use more personal pronouns except first person plural and third person plural.
			One explanation for the contrast between \emph{I} and \emph{we} is that
    		media outlets prefer statements about candidates themselves to unifying statements using \emph{we}.%
    		}
			& i, you, she, he & $\uparrow\uparrow\uparrow\uparrow$\\
			& &	they & \\
			& & we & $\downarrow\downarrow\downarrow\downarrow$\\
			\noalign{\smallskip}
			\hline

			\multirow{2}{\hsize}{Uncertainty/ subjectivity}
			& \multirow{2}{\hsize}{Hedging is a common way to express uncertainty \cite{lakoff1975hedges} and we use a dictionary from \citet{tan+lee:16tad}.
    		In the debate context, hedges may also represent subjectivity.
    		}
			& \multirow{2}{*}{hedges} & \multirow{2}{*}{$\uparrow\uparrow\uparrow\uparrow$}\\
			& & & \\
			\noalign{\smallskip}
			\hline

			\multirow{2}{\hsize}{Strong emphasis}
			& \multirow{2}{\hsize}{Superlatives represent the extreme form of an adjective or an adverb and can be used to put emphasis on a statement.
    		Surprisingly, \positive sentences do not use more superlatives.}
			& \multirow{2}{*}{superlatives} & \\
			& & & \\
			\noalign{\smallskip}
			\hline

			\multirow{2}{*}{Generality}
			& \multirow{2}{\hsize}{We count indefinite articles to measure generality. Our findings are consistent with \citet{Danescu-Niculescu-Mizil+Cheng+Kleinberg+Lee:12,Shahaf:2015:IJI:2783258.2783388,Tan:ProceedingsOfAcl:2014}.
			} 
			& \multirow{2}{*}{indef. articles} & \multirow{2}{*}{$\uparrow\uparrow\uparrow$}\\
			& & & \\
			\noalign{\smallskip}
			\hline

			\multirow{4}{\hsize}{Language model}
			& \multirow{4}{\hsize}{To capture surprise or conspicuousness, 
			we compute language model scores 
			based on NYT texts and 
    		part-of-speech (POS) tags in the WSJ portion of Penn Treebank.
    		However, the only significant feature 
    		is that \positive sentences are more similar to NYT texts in unigram usage. This finding is consistent with 
    		message sharing \cite{Tan:ProceedingsOfAcl:2014}
    		but is different from 
    		memorable movie quotes \cite{Danescu-Niculescu-Mizil+Cheng+Kleinberg+Lee:12}.} 
			& unigram & $\uparrow\uparrow$ \\
			& & bi-, trigram & \\
			& & POS \{1, 2, 3\}-gram & \\
			\noalign{\smallskip}
			\hline

			\multirow{3}{*}{Parallelism}
			& \multirow{3}{\hsize}{Using parallel sentence structure is a rhetorical technique, e.g, the first sentence in \tableref{tab:example} and ``I've never wilted in my life, and I've never wavered in my life''.
		    We use average longest common sequences between sub-sentences to measure it \cite{songlearning}.} 
			& \multirow{3}{*}{parallelism} & \multirow{3}{*}{$\uparrow$}\\
			& & & \\
			& & & \\
			\bottomrule
		\end{tabulary}
	\end{center}
	\caption{Testing results of sentence-alone features. Upward arrows
          indicate that \positive sentences have larger scores in that
          feature, while downward arrows suggest the other way around
          ($\uparrow\uparrow\uparrow\uparrow: p < 0.0001$,
          $\uparrow\uparrow\uparrow: p < 0.001$, $\uparrow\uparrow: p
          < 0.01$, $\uparrow: p < 0.05$, the same for downward arrows; $p$ refers to the $p$-value after the Bonferroni correction).} 
    \label{tab:sentence_features} 
\end{table*}

\subsection{Quantitative Features}
\label{sec:machine_features}

Building on the above human intuitions and existing studies, we
develop two sets of features: sentence-alone features, and
conversation-flow features that attempt to capture conversational
dynamics.
In this section, we use training data to identify important features that distinguish highlights from non-highlights, and compare the signals from data-driven methods with the factors from humans' free-form explanations.

In addition to these two sets of features, 
we will employ bag-of-words features as a strong baseline in \secref{sec:performance}, i.e., unigrams and bigrams that occur at least 5 times in the training set.

\para{Sentence-alone features.} 
We first examine features that do not rely on any contextual information in the debates and can be extracted from a sentence alone.
We evaluate whether \positive sentences are significantly different from \negative sentences in each feature.
Specifically, for each feature, we compute the feature values for both \positive and \negative sentences and
conduct one-sided paired $t$-tests with the Bonferroni correction \cite{bonferroni1936teoria}.
\tableref{tab:sentence_features} presents intuitions and theories for each feature set, including related work.
We 
present the full computational details in the appendix.

By comparing results in \tableref{tab:sentence_features} with previously discussed factors from human surveys, we find that the top factors in human surveys also tend to be statistically significant signals from data-driven methods, such as length (informative), and negative emotions (sensational).
But this is not always the case, e.g., positive emotions and strong emphasis were not statistically significant signals.
Meanwhile, signals such as personal pronouns, hedges, and language model features arise from data-driven methods, but humans may not pay as much attention to them.
A complete comparison between computational features and human factors
would require operationalizing controversiality, sensationalism, humor, etc.;
we leave this to future work.

\para{Conversation-flow features.}
Although only a handful of participants in our human experiment mentioned that context matters, %
conversational dynamics in the debates may contribute to the selection of highlights \cite{Zhang:ProceedingsOfNaacl:2016}.
We propose a novel set of conversation-flow features and indeed observe intriguing conversational dynamics around the highlights.

In order to capture the local context of a sentence ($s$),
we compare the sentence with its neighboring turns.
We use a window $w$ and denote the content words in the next $w$ turns by the same speaker as $\operatorname{Words}^{\mathrm{post}}_{\mathrm{self}}(w)$, the content words in the previous $w$ turns by the same speaker as $\operatorname{Words}^{\mathrm{prev}}_\mathrm{self}(w)$.
Similarly, we extract $\operatorname{Words}^{\mathrm{post}}_\mathrm{other}(w)$ and $\operatorname{Words}^{\mathrm{prev}}_\mathrm{other}(w)$ for other speakers.
We compute Jaccard similarity between 
the sentence ($\operatorname{Words}_s$) and 
its neighboring turns.
For instance,
$$
\operatorname{Jaccard}^{\mathrm{post}}_{\mathrm{other}}(5)=
\frac
{\lvert \operatorname{Words}_s \cap \operatorname{Words}^{\mathrm{post}}_{\mathrm{other}}(5) \rvert}
{\lvert \operatorname{Words}_s \cup \operatorname{Words}^{\mathrm{post}}_{\mathrm{other}}(5) \rvert}$$
\noindent measures the similarity between the 
sentence and 
the 5 turns by other speakers after the sentence.

\begin{figure}[t]
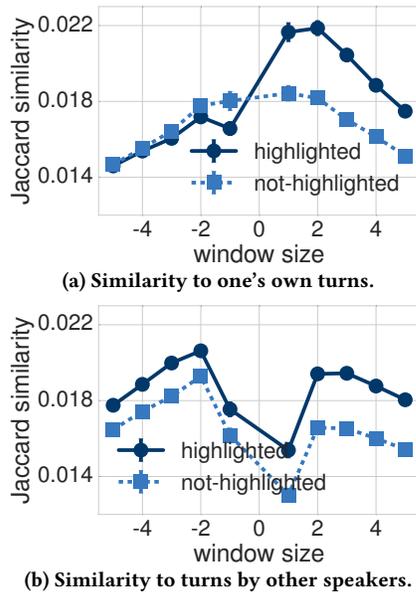

    \centering
    \begin{subfigure}[t]{0.32\textwidth}
        \addFigure{\textwidth}{quotability_feature/{jaccard_self_content_window_0.1}.pdf}
        \caption{Similarity to one's own turns.}
        \label{fig:simi_self}
    \end{subfigure}
    \begin{subfigure}[t]{0.32\textwidth}
        \addFigure{\textwidth}{quotability_feature/{jaccard_other_content_window_0.1}.pdf}
        \caption{Similarity to turns by other speakers.}
        \label{fig:simi_other}
    \end{subfigure}
    \caption{%
    \figref{fig:simi_self} and \figref{fig:simi_other} present conversation-flow features that are based on Jaccard similarity between a sentence and its neighboring turns (negative windows for previous turns, positive windows for later turns, error bars are tiny).
    In \figref{fig:simi_self}, a kink exists around 0 
    in similarity to turns by the same speaker, while in \figref{fig:simi_other}, \positive sentences are consistently more similar to turns by other speakers.
    \label{fig:flow_features}}
\end{figure}

\begin{itemize}[leftmargin=*]
	\item A kink exists in similarity to turns by the same speaker (\figref{fig:simi_self}).
	\Positive and \negative sentences present the same level of similarity
	to turns by the same speaker until 
	the last turn before the sentence.
	An interesting kink shows up around the sentence:
	\positive sentences are less similar to the turn immediately before
	but are more similar to turns after.
    We take this as a sign that ``changepoints'' in a
	monologue, where the speaker shifts in topic or style, are more likely
	to be quoted.
	\item \Positive sentences are more similar to neighboring turns by other speakers (\figref{fig:simi_other}).
	Regarding the overall trend, 
	both for \positive and \negative sentences,
	the similarity is smaller 
	for the immediate neighboring turns ($w$ is 1 or -1) than when more turns are included.
	This is because moderators often speak right before and
	right after candidates, and 
	moderators speak distinctly from candidates, in terms of words (due to different communicative goals).
\end{itemize}

\subsection{Prediction Performance}
\label{sec:performance}

Finally, we study to what extent media choices can be predicted only from textual factors by examining classification performance on the held-out set.
We also investigate the difference between general debates and primary
debates.

\begin{figure*}
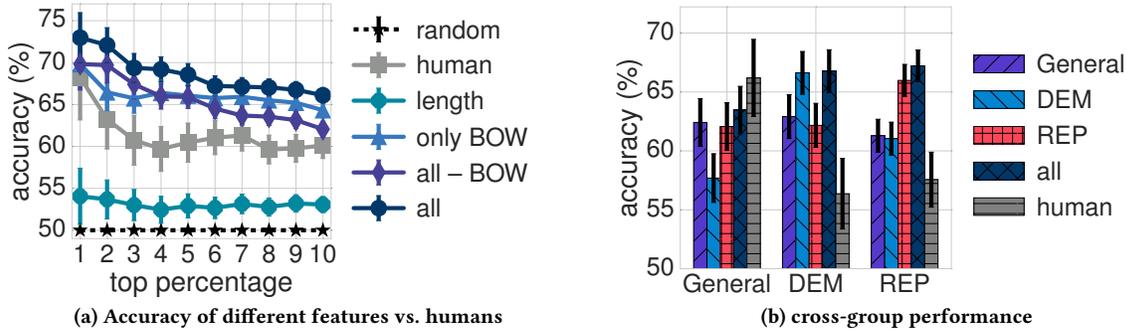

    \centering
    \begin{subfigure}[t]{0.45\textwidth}
        \centering
        \addFigure{\textwidth}{performance/{heldout_subset_0.1}.pdf}
        \caption{Accuracy of different features vs. humans}
        \label{fig:heldout_10}
    \end{subfigure}
    \begin{subfigure}[t]{0.45\textwidth}
        \centering
        \addFigure{\textwidth}{performance/{all_0.1_debate_group_with_all}.pdf}
        \caption{cross-group performance}
        \label{fig:cross_debate_type}
    \end{subfigure}
    \caption{Classification accuracy. 
    In \figref{fig:heldout_10}, each point measures the accuracy of 
    a feature set (indicated by the color) on the subset of held-out data where the quote count of the \positive sentence in a pair is among the top $x$\% quoted sentences in the debate.
    In other words, smaller $x$ values correspond to ``easier'' pairs where the \positive sentence is more prominent.
    $y$ values for $x=10$ give the accuracies on the full held-out
    data (all, 66.0\% vs. human, 60.1\%, $p= 0.0008$).
    \figref{fig:cross_debate_type} illustrates the accuracy when we apply the classifier trained on a debate type to another debate type. 
    Different colors represent the debate type of the training data, 
    and the $x$-axis represents the debate type of the test data.
    Note that ``human'' reports the accuracy of human predictions on
    the corresponding debate type in the test data, and is not a
    function of the training set.
    }
    \label{fig:performance}
\end{figure*}

\para{Overall prediction accuracy (\figref{fig:heldout_10}).}
Using only textual factors, our classifier achieves an accuracy of 66\% on the held-out set for $t=10$.
The accuracy of both machines and humans increases for ``easier'' pairs,
the ones in which \positive sentences were quoted more frequently.
This trend confirms that meaningful signals exist in the wording.
The accuracy of machines (``all'') is always above humans and the difference is statistically significant.

The bag-of-words (BOW) model already outperforms humans in this task.
In comparison with the features that we propose in
\secref{sec:machine_features} (\notbow),
although \notbow has far fewer features, it 
yields a similar accuracy to BOW.
\notbow 
works relatively well when \positive sentences in a pair 
were quoted more frequently and when there are fewer training instances, while BOW has an advantage 
when the \positive sentence is closer to the 10\% threshold (right side of \figref{fig:heldout_10}).
Combining all features (including BOW; ``all'') always leads to the best accuracy.

\emph{Note that the accuracies of machines and humans are not meant to be compared head-to-head, since machines rely on training data to identify the useful signals, while humans depend on their daily (potentially biased) media exposure.}
Instead, we view this accuracy gap as evidence suggesting that some signals in the wording are hard for humans to identify.
This also points to the potential to inform the public with the help of machines.

\para{Differences across debate types (\figref{fig:cross_debate_type}).}
As primary debates have more candidates and receive less coverage than general debates,
the news media may employ different criteria to select \highlights.
To explore the differences,
we train classifiers on subsets of the training data from primary debates and test on different types of debate.
Differences indeed exist in how wording affects media selection: 
the classifiers do not usually perform well when tested on other debate types, except from Republican primary debates to general debates.
In fact, using all training instances does not improve 
over using the pairs only from the matching debate type, despite the
latter's use of fewer training instances.\footnote{A more comparable setup is to subsample training instances to match the size in a particular debate type.
	Doing this, training only using the matching debate type
        (known as ``in-domain'') always outperforms using all debate types.
	}

\para{Machines outperform humans only in primary debates.} 
Human accuracy is much better in general debates than in primary debates.
In fact, the advantage of our classifiers (``all'') in \figref{fig:heldout_10} mainly comes from primary debates.
The reason may be that general debates receive more attention and more coverage,
humans are thus more likely to remember what was selected as \highlights; indeed, ``memory, past exposure'' was an important factor in the surveys.
If this is the case, humans may have an even more limited understanding of media choices
had there been no influence from previous exposure.
We also observe that humans perform better in general debates after 2000 than in those before 2000.

\section{Media Preferences over Time}
\label{sec:media_preferences}

Beyond the effect of wording,
a media outlet's own preferences can potentially impact how \highlights are
selected.
In fact, media polarization 
has attracted significant interest from both researchers and the public
\cite{Baum:PoliticalCommunication:2008,iyengar2009red}.
We take advantage of the longitudinal nature of our dataset, and
evaluate the extent of media fragmentation over time.
Building on the intuition that outlets are similar if they quote the same sentences with similar sentiments, we employ two approaches to quantify the fragmentation level.
We first consider the existing two-party structure in the U.S. and evaluate %
whether the media quote both parties ``evenly''.
Second, we borrow concepts from the clustering literature and examine the overall similarity between media outlets beyond the partisan assumption.
%
%
%
%
%
%

\subsection{Bipartisan Coverage}

\begin{figure}[t]
\centering
\includegraphics[clip, trim=0.cm 0.2cm 0.0cm 0.2cm,width=0.47\textwidth]{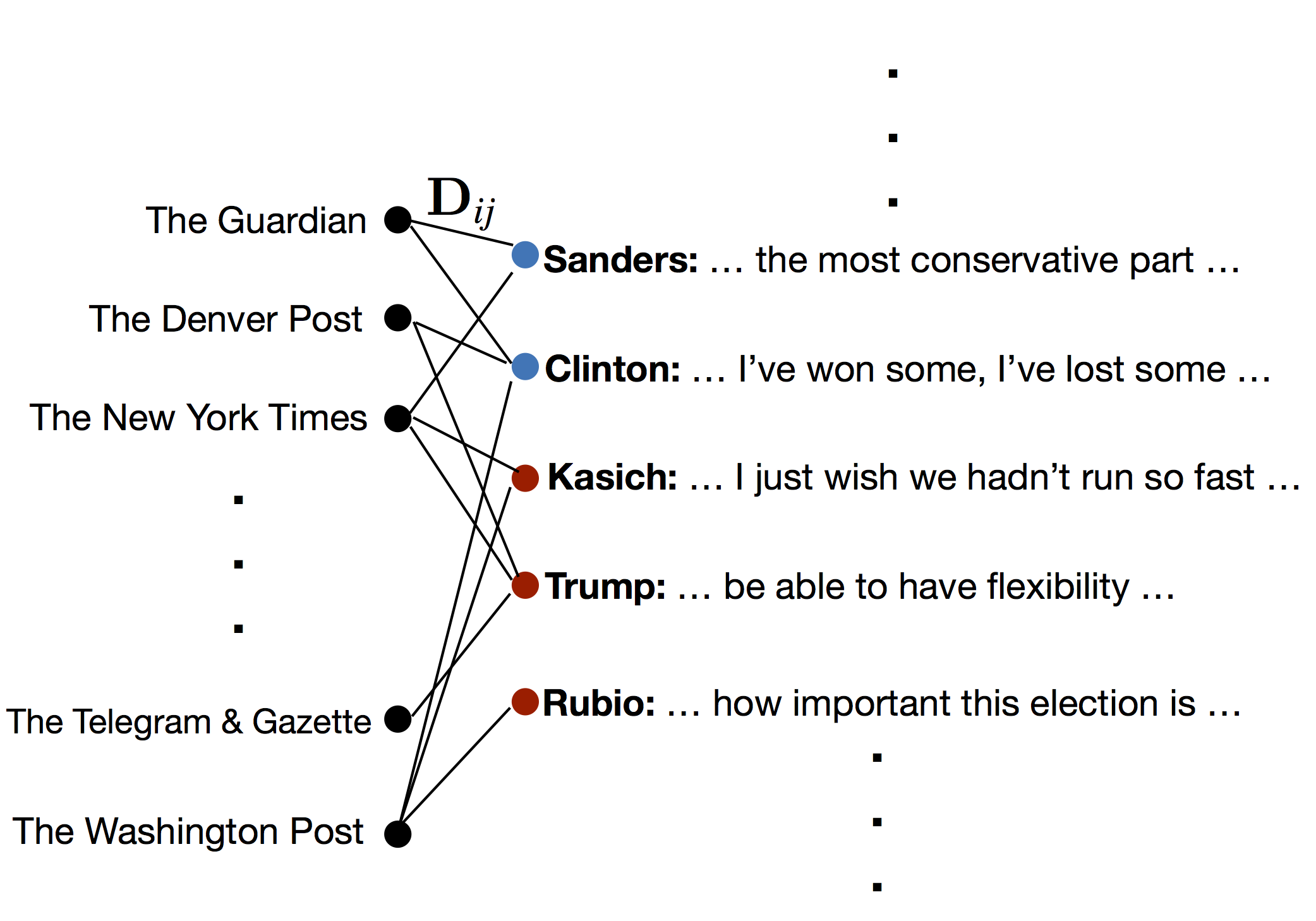}
\vspace{-.35cm}
\caption{Bipartite graph between media outlets and highlights
  from presidential debates. The edges are sampled from presidential
  debates in 2016.
}
\label{fig:bipartite_example}
\end{figure}

Because 
U.S. presidential elections 
typically involve two major parties,
we first take advantage of this two-party structure
and 
evaluate fragmentation by how much ``Democratic-leaning'' media outlets quote Republican candidates and vice versa.

\para{Bipartite graph representation.}
A natural representation of quoting patterns is a bipartite graph between outlets and sentences, where an edge between media outlet $i$ and candidate sentence $j$ indicates that $i$ quotes $j$ (e.g., \figref{fig:bipartite_example}).
This graph can be represented using a media-sentence matrix $\msmatrix \in
\mathbb{R}^{M \times S}$, where each row represents a media outlet
($M$ is the number of 
outlets) and each column represents
a sentence from a candidate ($S$ is the number of sentences).
To obtain $\msmatrix_{ij}$, we use three methods to account for both the frequency and the sentiment of a media outlet quoting a candidate utterance:
{\em a.~count} ($\msmatrix_{ij}$ is the number of times that sentence $j$ was quoted in outlet $i$);
{\em b.~positive context} ($\msmatrix_{ij}$ is the number of positive words in the 30 words around each quote of sentence $j$ in outlet $i$);
{\em c.~negative context} (similar to positive context but counting negative words). 
We normalize each row so that the $\ell_2$-norm is 1 and remove media outlets that used fewer than 10 quotes in an election.

\begin{figure*}[t]
    \centering
    \begin{subfigure}[t]{0.3\textwidth}
        \addFigure{\textwidth}{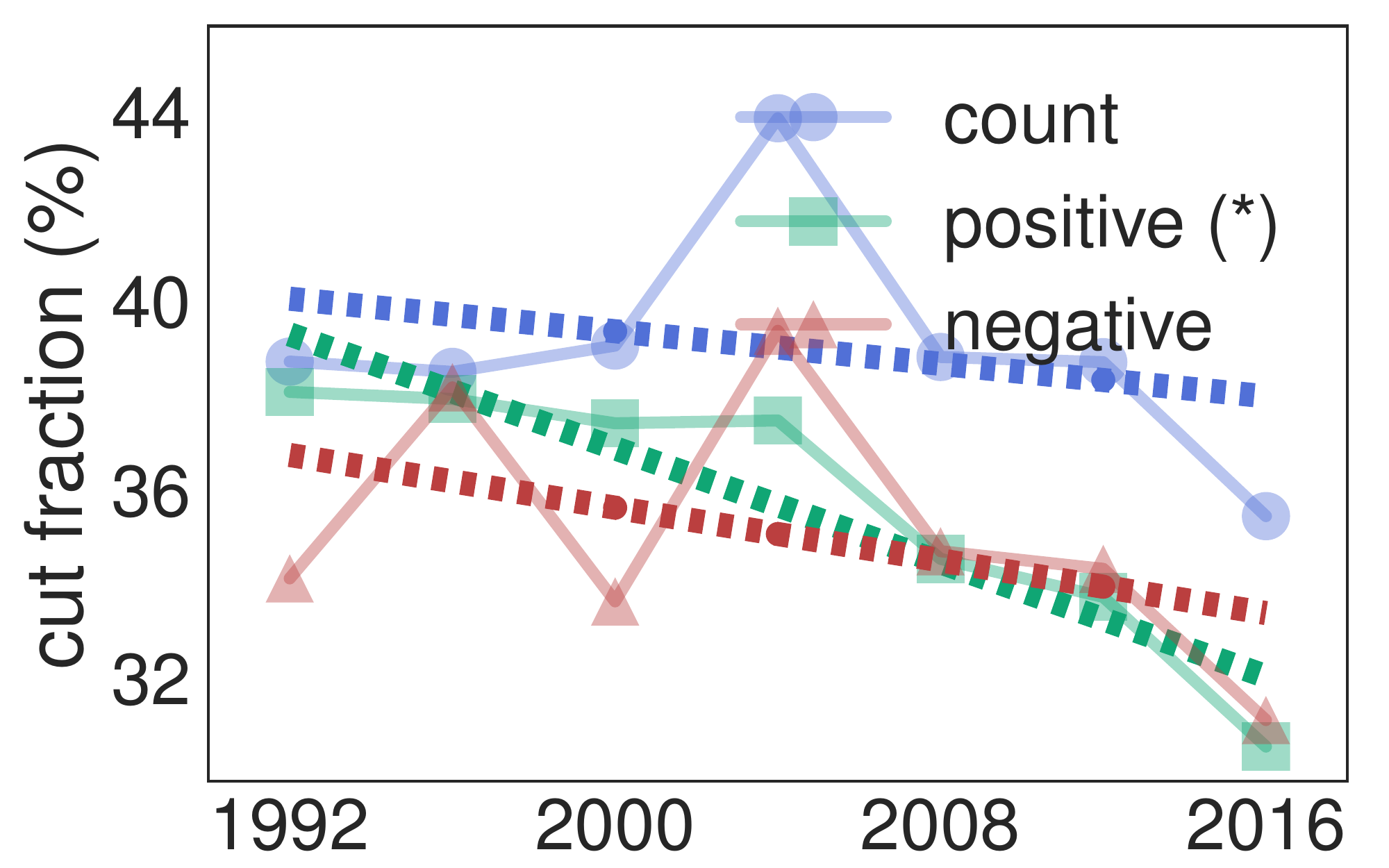}
        \caption{The fraction of weights in the min-cut.}
        \label{fig:cut_fractions}
    \end{subfigure}
    \hfill
    \begin{subfigure}[t]{0.3\textwidth}
        \addFigure{\textwidth}{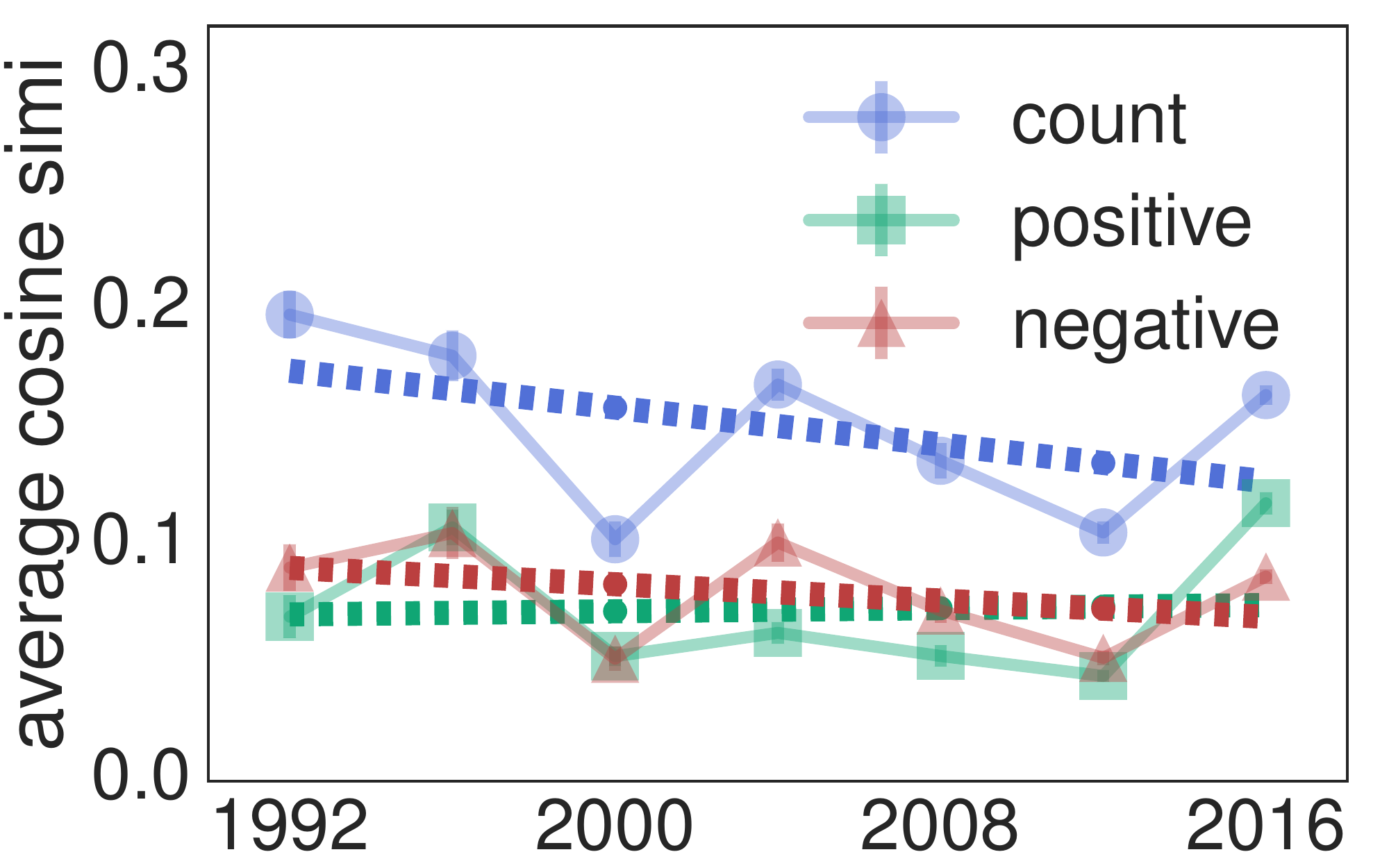}
        \caption{Average pairwise similarity across all media outlets.}
        \label{fig:bias_global_dist}
    \end{subfigure}
    \hfill
    \begin{subfigure}[t]{0.3\textwidth}
        \addFigure{\textwidth}{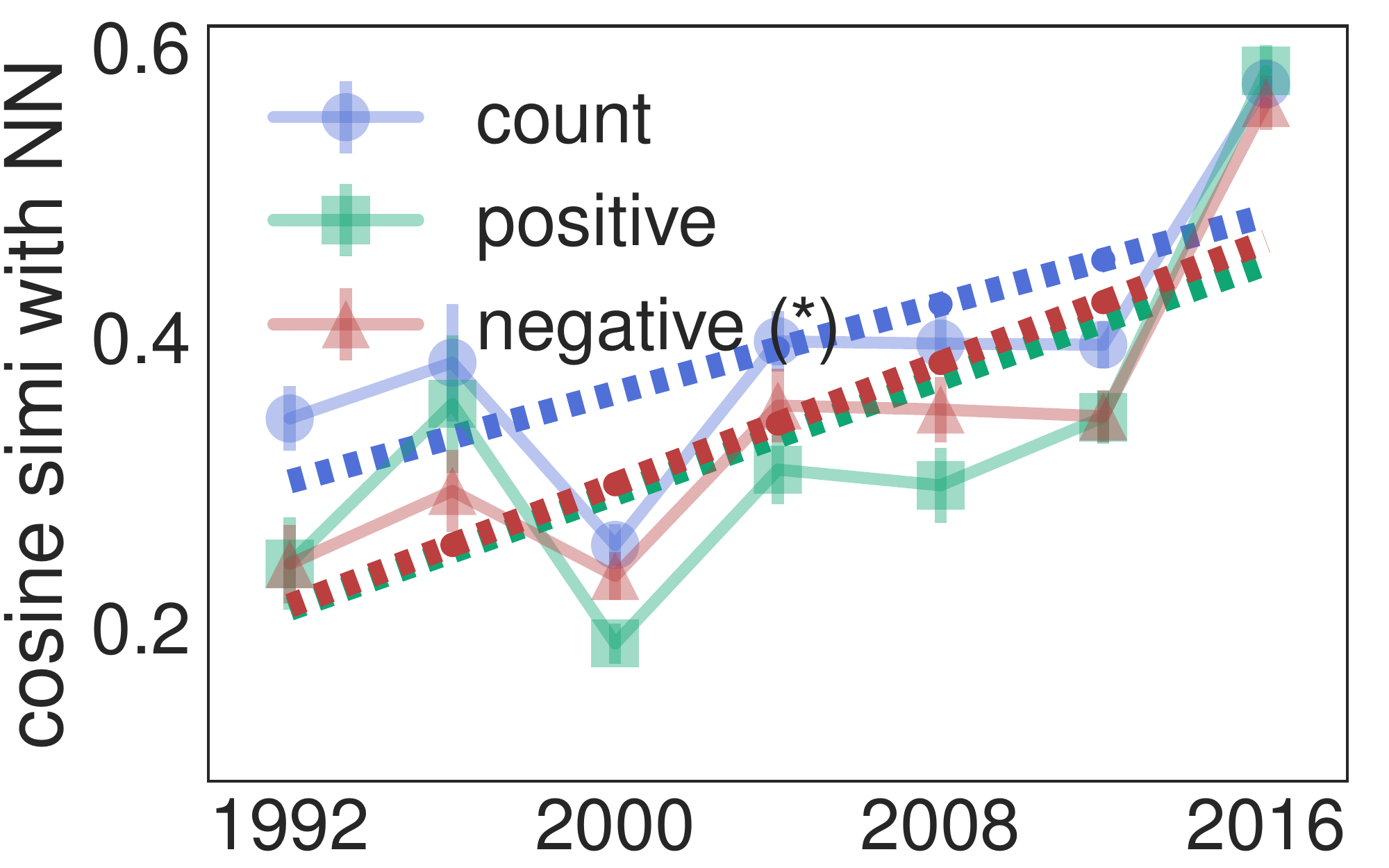}
        \caption{Average similarity with 3 nearest neighbors.}
        \label{fig:bias_local_dist}
    \end{subfigure}
    \caption{\figref{fig:cut_fractions} 
    estimates whether the outlets quote evenly across two parties and
    shows a declining trend over time.
    \figref{fig:bias_global_dist} gives the global average pairwise similarity and there are no clear trends,
    while \figref{fig:bias_local_dist} shows that ``local similarity'' (average similarity with top nearest neighbors) increases over time.
    There are not enough media outlets with at least 10 quotes to
    compute meaningful results in the 80s, so we exclude those years.
    In all figures, different colors represent different ways to represent media outlets with their quoting patterns.
    \label{fig:bias}}
\end{figure*}

\para{Using min-cut to identify bipartisan coverage.} 
We focus on news coverage of general debates, because there has
always been one presidential candidate and one vice-presidential
candidate from the Democratic party and the Republican party during
the past three decades.\footnote{We ignore all independent candidates in this analysis.}
We thus build matrices for the bipartite graphs based on general debates in each presidential election.

To the extent that outlets ``lean''
one way or the other, we expect them to quote one party or the other
more (or to positively quote one party more, or to negatively quote one
party more).
In the extreme case, a subset of media outlets might only quote the Republican candidates and the rest only quote the Democratic candidates, in other words, there exists no bipartisan coverage.
These intuitions align with the idea of using min-cut to identify bipartisan coverage.
If we apply the min-cut algorithm to separate sentences from Democratic
candidates and sentences from Republican candidates in the bipartite graph
\cite{stoer1997simple}, 
then the extreme case where no bipartisan coverage exists leads to a min-cut of 0.
Conversely, the more costly the min-cut, the more
entangled the two sides are. 
\para{Decreasing bipartisan coverage (\figref{fig:cut_fractions}).}
We thus compute the fraction of weights in the min-cut and smaller values in this statistic
indicate that 
media outlets quote largely from one of the two sides and little from
the other.\footnote{Note that media outlets that preferentially quote
  Democratic candidates may not support the Democratic party, because
  a quote can be presented in a negative light.
  We thus also use sentiment information in the context of quotes to populate $\msmatrix$.
  }
\figref{fig:cut_fractions} shows that 
cross-cutting coverage in the min-cut is declining over time, under
all three definitions.
It is worth noting that the fraction of weights in the min-cut is not small (about 40\%, upper bounded by 50\%) despite the declining trend, which suggests that media outlets tend to at least cover both sides.
The fact that there exists less cross-cutting coverage in sentiment (both positive and negative) than in counts indicates that although media outlets quote both sides, the sentiment differs.\footnote{In terms of the partition resulted from the min-cut, in most years, the majority of the media outlets are in the partition associated with Republican candidate sentences, at least consistent with a recent report in 2016 \cite{patterson2016news}.
This is especially true for big media outlets such as the \emph{New York Times} and \emph{Washington Post}. A notable exception is that the \emph{Washington Post} is in the Democratic partition for positive contexts in 2008.}

\subsection{Beyond Partisan Assumptions}

An alternative way to estimate fragmentation is to study how well media outlets cluster together without partisan assumptions.
We build matrices based on both primary debates and general debates in each presidential election.
We use each row $\msmatrix_i$ to represent media outlet $i$ and investigate the quality of clustering between media outlets.

Clustering purity is typically evaluated 
at two levels: in a pure clustering,
inter-cluster distances 
are large and intra-cluster distances are small \cite{rousseeuw1987silhouettes}.
When we use the Silhouette score (a measure of purity) to identify the
optimal number of clusters in a $K$-means clustering of media representations in $\msmatrix$,
%
%
%
the optimal number is close to the total number of media outlets, $M$, which suggests that there are many isolated small clusters.
We thus examine fragmentation at the above two levels by considering each media outlet as a singleton:
whether media outlets have become less similar 
to each other overall (analogous to 
inter-cluster distances); and
whether media outlets have become more similar to their nearest neighbors
(analogous to 
intra-cluster distances).
We use cosine similarity to measure the similarity between a pair of media outlets.

\para{No clear trends in ``inter-cluster'' similarity (\figref{fig:bias_global_dist}).}
To evaluate whether the news media become less similar to each other, we calculate the global mean of all pairwise similarities.
A decreasing 
global mean would indicate fragmentation at the
global level, but we do not observe consistent trends or any
statistically significant correlation with time (\figref{fig:bias_global_dist}).
The similarity in positive context and negative context is always
smaller than the similarity in usage frequency.
This again suggests that, although different media outlets may quote the same sentences, they present different opinions around the quotes. 
\para{Increasing ``intra-cluster'' similarity (\figref{fig:bias_local_dist}).}
To capture ``local'' similarity that is analogous to intra-cluster distances,
we propose a statistic that 
measures the average similarity between a media outlet and
its $K$ nearest neighbors.
We refer to this as {\em local similarity}.
A growing local similarity suggests increasing tightness at the local level.
We observe consistent increasing trends across three definitions, and
this observation is robust with choices of $K$.
This observation is related to the fact that there are increasingly many media outlets over time and it is thus more likely for a media outlet to have a close nearest neighbor.
However, this hypothesis is insufficient to explain our observations,
because  it also suggests that ``inter-cluster'' similarity should increase, which does not hold.

\para{Discussion.}
Our observations are derived from a three-decade dataset and are consistent with past work on polarization and partisan selective exposure \cite{Baum:PoliticalCommunication:2008,Stroud:JournalOfCommunication:2010},
but the reasons behind the decreasing bipartisan coverage and the increasing %
local similarity
require further investigation.

Our results are certainly limited by the relatively short history of presidential debates.
It is also important to note that our study does not take into account the
influence among media outlets themselves
\cite{Golan:JournalismStudies:2007,boyle:2001,kim+barnett:1996}. 
For instance, \citet{Golan:JournalismStudies:2007} shows a correlation between the morning \emph{New York Times} and three evening television news programs.
Further studies regarding the diffusion in media selection of highlights
can shed more light on our observations.

\section{Related Work}
\label{sec:related}

We have discussed the most relevant studies throughout the paper.
Here we discuss three additional strands of related work.

\para{The effect of post-debate coverage on public opinion.} 
Studies have shown that media choices about coverage can 
have serious consequences \cite{brubaker2009effect,Fridkin01012008,Hillygus:AmericanJournalOfPoliticalScience:2003,hwang2007applying,Tsfati01072003,patterson2016press,gross+etal:17}.
For instance, \citet{Fridkin01012008} show that in the 2004 U.S. election, 
citizens who only read the news coverage 
rated Kerry more negatively compared to those who watched the debate firsthand, because media outlets highlighted the moment of Kerry outing Cheney's lesbian daughter, although this moment did not catch much attention from the live audience.
\citet{boydstun2014real} develop a mobile app to collect real-time feedback for presidential debates.
\citet{patterson2016press} discuss the media's critical tendency and its partisan consequences in the U.S.

\para{Influences between the media and politicians.}
Although our work focuses on 
media selection of highlights, 
politicians often behave based on their beliefs about media preferences, which suggests complex dynamics between the media and politicians \cite{KarenCallaghan:PoliticalCommunication:2010,blumler1999third,grand2015emergence,clayman2002news,prat2011political}.
For instance, \citet{blumler1999third} discuss 
politicians' increasing adaptation to different news values and formats 
in the presence of media abundance.
Also relevant is research on 
the influence of politicians on the media, including agenda-setting, rhetorical positioning, and framing \cite{mccombs1972agenda,Chong:JournalOfCommunication:2007,Entman:JournalOfCommunication:1993,sim:2013:emnlp,Wells:PoliticalCommunication:2016}.

\para{Power dynamics in debates and other types of coverage.}
Studies have shown that language use and topic control in debates can reflect influence between candidates and indicate power dynamics
\cite{Nguyen:2012:SHN:2390524.2390536,Nguyen:2014:MTC:2629710.2629739,prabhakaran2014staying}.
More recently, 
social media have also become an important channel to monitor  public
opinion on debates in real time \cite{broersma2012social,diakopoulos2010characterizing}
and potentially change news media coverage. 

\section{Conclusion}
\label{sec:conclusion}

In this paper, 
we conduct the first systematic study on 
media selection of \highlights from presidential debates, using a three-decade dataset.
We introduce a computational framework that controls for the speaker and the debate situation to study the effect of textual factors.
First, we find that 
media choices are not obvious to 
Mechanical Turk workers,
suggesting that the public may 
have a limited understanding of how the news media choose \highlights in news coverage.
Second, although machines
and humans achieve similar accuracy in general debates, machines
significantly outperform humans in predicting media-chosen \highlights in primary debates.
Our findings indicate that there exist signals in the textual
information that untrained humans do not find salient.
In particular, 
highlights are locally distinct from the speaker's previous turn, but are later echoed more by both the speaker and other participants.
We further demonstrate a declining trend of bipartisan coverage using macro quoting patterns and analyze the quality of clustering between media outlets without partisan assumptions.

The news media play an important role in connecting the public and politicians. 
Our work indicates that the public may not understand what factors matter in media choices.
Quantitative studies in this domain can complement qualitative theories to improve the understanding of political and media communication for both scholars and the public.

    \para{Acknowledgments.} We thank Yejin Choi, Aaron Jaech, Luke Zettlemoyer,
    anonymous reviewers, and all members of Noah's ARK for helpful comments and discussions. This research was made possible by a University of Washington Innovation Award.
    
    \appendix

\section{Appendix}

%

%
%
%
%

%
%
%

%
%
%
%
%

%
%
%
%
%
%
%

%
%
%
%
%
%
%
%
%

\subsection{Amazon Mechanical Turk Labeling Details}

\para{Detailed task and filtering criteria.}
For each threshold $t \in \{2, 4, \\ 6, 8, 10\}$,
we randomly sampled 200 pairs from the corresponding heldout set, in total 1,000 pairs.
In our Amazon Mechanical Turk experiments, we required participants to be from the United States and have done at least 10 HITs with at least 97\% acceptance rate.
We requested each participant to label 25 pairs in an assignment (5 for each threshold) and each participant can only finish one assignment.
We paid \$1.00 (\$0.04 a pair) for an assignment
and gave \$0.02 bonus for each correct guess to incentivize participants to try their best.
We removed participants who did not answer survey questions in good faith and spent too short time on the task.
To compensate for such filtering, we added assignments on Amazon Mechanical Turk until we had 200 valid assignments.
In the end, we gathered 5,000 labels in total (5 labels per pair).

\para{Survey questions.}
The following questions were asked after participants finished labeling:
\begin{itemize}[leftmargin=*]
    \item Some people follow  presidential debates most of the time, while others aren't that interested. How often would you say you have watched presidential debates from 1980 to 2016? (Never, 1-5 times, 5-10 times, more than 10 times)
    \item Generally speaking, when it comes to political parties in the U.S., how would you describe yourself? (Democrat, Independent close to Democrat, Independent (close to neither), Independent close to Republican, Republican, Other)
    \item Explain what factors influence your decision on which sentence was quoted more, simple comments such as several adjectives or nouns (e.g., issues, surprise) can help our research. (free-form response)
    \item If you have any comment about our task, please give us your feedback. (free-form response)
\end{itemize}

There are no clear trends in human performances regarding levels of experience or ideology.
Most participants gave very positive feedback about our task, e.g., %
``Very fun to read some of the old transcripts and think about those conversations--thank you!''.

\subsection{Sentence-alone Feature Definitions and Testing Results}

\para{Detailed definitions.} The following list is aligned with features in 
Table 2.
\begin{itemize}[noitemsep,topsep=1pt,leftmargin=*]
    \item Length is measured by the number of words in a sentence.
    \item The lexicons of positive words and negative words come from the corresponding category in LIWC \cite{pennebaker2001linguistic}.
    \item Negations and negative conjunctions. Negations include ``n't'', ``not'', ``no'', ``cannot''. Negative conjunctions include ``although'', ``atho'', ``but'', ``nor'', ``whereas'', ``while'', ``though'', ``however'', ``otherwise'', ``tho'' and ``unless'.'
    \item Personal pronouns. The list of definitions come from LIWC \cite{pennebaker2001linguistic}.
    \item {Hedges.} We use a list of regular expressions from \citet{tan+lee:16tad}.
    \item {Superlatives.} We get a list of candidate lexicons by matching words with the regular expression ``[a-zA-Z]+est'', and then manual filter false positives. We also include ``most'', ``least'' and ``worst''.
    \item {Indefinite articles.} ``a'' and ``an''. 
    \item {Language model features}. 
    We use each sentence's likelihood w.r.t. both lexical level and part-of-speech level language models trained on newswire data.
    Specifically, we train 1, 2, and 3-gram lexical level language models on 
    the NYT corpus\footnote{LDC number: LDC2008T19. \url{https://catalog.ldc.upenn.edu/ldc2008t19}}, 
    and we use the WSJ portion of Penn Treebank\footnote{LDC number: LDC99T42. \url{https://catalog.ldc.upenn.edu/ldc99t42}} to train POS-level models.
	Our implementation is based on SRILM \cite{andreas2011srilm}.
    \item {Parallelism}. 
    We measure parallelism using the average length of the longest common sequences between sub-sentences following \citet{songlearning}.
    The basic unit in the longest common sequence is a word.
    Sub-sentences are split by ``;'' or ``,''.
\end{itemize}

\para{Feature testing procedure and results.}
For each feature, we use a one-sided paired $t$-test to test whether, on our training pairs, our scoring function for that feature is larger in the \positive sentences than in the \negative sentences. 
Given that we did 20 tests in total, there is a risk of obtaining false positives due to multiple testing. 
To account for this, we only report significant results after the
Bonferroni correction \cite{bonferroni1936teoria}, i.e., we multiply
each $p$-value by 20 and see whether it is smaller than (for
example) 0.05.

	\bibliographystyle{ACM-Reference-Format}
  \balance
  \bibliography{ref}


\begin{thebibliography}{58}


\ifx \showCODEN    \undefined \def \showCODEN     #1{\unskip}     \fi
\ifx \showDOI      \undefined \def \showDOI       #1{#1}\fi
\ifx \showISBNx    \undefined \def \showISBNx     #1{\unskip}     \fi
\ifx \showISBNxiii \undefined \def \showISBNxiii  #1{\unskip}     \fi
\ifx \showISSN     \undefined \def \showISSN      #1{\unskip}     \fi
\ifx \showLCCN     \undefined \def \showLCCN      #1{\unskip}     \fi
\ifx \shownote     \undefined \def \shownote      #1{#1}          \fi
\ifx \showarticletitle \undefined \def \showarticletitle #1{#1}   \fi
\ifx \showURL      \undefined \def \showURL       {\relax}        \fi
\providecommand\bibfield[2]{#2}
\providecommand\bibinfo[2]{#2}
\providecommand\natexlab[1]{#1}
\providecommand\showeprint[2][]{arXiv:#2}

\bibitem[\protect\citeauthoryear{Atkinson}{Atkinson}{1984}]%
        {atkinson1984our}
\bibfield{author}{\bibinfo{person}{Max Atkinson}.}
  \bibinfo{year}{1984}\natexlab{}.
\newblock \bibinfo{booktitle}{\emph{Our Masters' Voices: The Language and Body
  Language of Politics}}.
\newblock \bibinfo{publisher}{Psychology Press}.
\newblock


\bibitem[\protect\citeauthoryear{Baum and Groeling}{Baum and Groeling}{2008}]%
        {Baum:PoliticalCommunication:2008}
\bibfield{author}{\bibinfo{person}{Matthew~A. Baum} {and} \bibinfo{person}{Tim
  Groeling}.} \bibinfo{year}{2008}\natexlab{}.
\newblock \showarticletitle{New Media and the Polarization of American
  Political Discourse}.
\newblock \bibinfo{journal}{\emph{Political Communication}}
  \bibinfo{volume}{25}, \bibinfo{number}{4} (\bibinfo{year}{2008}),
  \bibinfo{pages}{345--365}.
\newblock


\bibitem[\protect\citeauthoryear{Blumler and Kavanagh}{Blumler and
  Kavanagh}{1999}]%
        {blumler1999third}
\bibfield{author}{\bibinfo{person}{Jay~G. Blumler} {and}
  \bibinfo{person}{Dennis Kavanagh}.} \bibinfo{year}{1999}\natexlab{}.
\newblock \showarticletitle{The Third Age of Political Communication:
  Influences and Features}.
\newblock \bibinfo{journal}{\emph{Political communication}}
  \bibinfo{volume}{16}, \bibinfo{number}{3} (\bibinfo{year}{1999}),
  \bibinfo{pages}{209--230}.
\newblock


\bibitem[\protect\citeauthoryear{Bonferroni}{Bonferroni}{1936}]%
        {bonferroni1936teoria}
\bibfield{author}{\bibinfo{person}{Carlo~E. Bonferroni}.}
  \bibinfo{year}{1936}\natexlab{}.
\newblock \bibinfo{booktitle}{\emph{Teoria statistica delle classi e calcolo
  delle probabilita}}.
\newblock \bibinfo{publisher}{Libreria internazionale Seeber}.
\newblock


\bibitem[\protect\citeauthoryear{Boydstun, Glazier, Pietryka, and
  Resnik}{Boydstun et~al\mbox{.}}{2014}]%
        {boydstun2014real}
\bibfield{author}{\bibinfo{person}{Amber~E. Boydstun},
  \bibinfo{person}{Rebecca~A. Glazier}, \bibinfo{person}{Matthew~T. Pietryka},
  {and} \bibinfo{person}{Philip Resnik}.} \bibinfo{year}{2014}\natexlab{}.
\newblock \showarticletitle{{Real-time Reactions to a 2012 Presidential Debate:
  A Method for Understanding Which Messages Matter}}.
\newblock \bibinfo{journal}{\emph{Public Opinion Quarterly}}
  \bibinfo{volume}{78}, \bibinfo{number}{S1} (\bibinfo{year}{2014}),
  \bibinfo{pages}{330--343}.
\newblock


\bibitem[\protect\citeauthoryear{Boyle}{Boyle}{2001}]%
        {boyle:2001}
\bibfield{author}{\bibinfo{person}{Thomas~P. Boyle}.}
  \bibinfo{year}{2001}\natexlab{}.
\newblock \showarticletitle{Intermedia Agenda Setting in the 1996 Presidential
  Election}.
\newblock \bibinfo{journal}{\emph{Journalism \& Mass Communication Quarterly}}
  \bibinfo{volume}{78}, \bibinfo{number}{1} (\bibinfo{year}{2001}),
  \bibinfo{pages}{26--44}.
\newblock


\bibitem[\protect\citeauthoryear{Broersma and Graham}{Broersma and
  Graham}{2012}]%
        {broersma2012social}
\bibfield{author}{\bibinfo{person}{Marcel Broersma} {and} \bibinfo{person}{Todd
  Graham}.} \bibinfo{year}{2012}\natexlab{}.
\newblock \showarticletitle{Social Media as Beat: Tweets as a News Source
  During the 2010 British and Dutch elections}.
\newblock \bibinfo{journal}{\emph{Journalism Practice}} \bibinfo{volume}{6},
  \bibinfo{number}{3} (\bibinfo{year}{2012}), \bibinfo{pages}{403--419}.
\newblock


\bibitem[\protect\citeauthoryear{Brubaker and Hanson}{Brubaker and
  Hanson}{2009}]%
        {brubaker2009effect}
\bibfield{author}{\bibinfo{person}{Jennifer Brubaker} {and}
  \bibinfo{person}{Gary Hanson}.} \bibinfo{year}{2009}\natexlab{}.
\newblock \showarticletitle{The Effect of Fox News and CNN's Postdebate
  Commentator Analysis on Viewers' Perceptions of Presidential Candidate
  Performance}.
\newblock \bibinfo{journal}{\emph{Southern Communication Journal}}
  \bibinfo{volume}{74}, \bibinfo{number}{4} (\bibinfo{year}{2009}),
  \bibinfo{pages}{339--351}.
\newblock


\bibitem[\protect\citeauthoryear{Callaghan and Schnell}{Callaghan and
  Schnell}{2010}]%
        {KarenCallaghan:PoliticalCommunication:2010}
\bibfield{author}{\bibinfo{person}{Karen Callaghan} {and}
  \bibinfo{person}{Frauke Schnell}.} \bibinfo{year}{2010}\natexlab{}.
\newblock \showarticletitle{Assessing the Democratic Debate: How the News Media
  Frame Elite Policy Discourse}.
\newblock \bibinfo{journal}{\emph{Political Communication}}
  \bibinfo{volume}{18}, \bibinfo{number}{2} (\bibinfo{year}{2010}),
  \bibinfo{pages}{183--213}.
\newblock


\bibitem[\protect\citeauthoryear{Chong and Druckman}{Chong and
  Druckman}{2007}]%
        {Chong:JournalOfCommunication:2007}
\bibfield{author}{\bibinfo{person}{Dennis Chong} {and}
  \bibinfo{person}{James~N. Druckman}.} \bibinfo{year}{2007}\natexlab{}.
\newblock \showarticletitle{A Theory of Framing and Opinion Formation in
  Competitive Elite Environments}.
\newblock \bibinfo{journal}{\emph{Journal of Communication}}
  \bibinfo{volume}{57}, \bibinfo{number}{1} (\bibinfo{year}{2007}),
  \bibinfo{pages}{99--118}.
\newblock


\bibitem[\protect\citeauthoryear{Clayman}{Clayman}{1995}]%
        {clayman1995defining}
\bibfield{author}{\bibinfo{person}{Steven~E. Clayman}.}
  \bibinfo{year}{1995}\natexlab{}.
\newblock \showarticletitle{Defining Moments, Presidential Debates, and the
  Dynamics of Quotability}.
\newblock \bibinfo{journal}{\emph{Journal of Communication}}
  \bibinfo{volume}{45}, \bibinfo{number}{3} (\bibinfo{year}{1995}),
  \bibinfo{pages}{118--147}.
\newblock


\bibitem[\protect\citeauthoryear{Clayman and Heritage}{Clayman and
  Heritage}{2002}]%
        {clayman2002news}
\bibfield{author}{\bibinfo{person}{Steven~E. Clayman} {and}
  \bibinfo{person}{John Heritage}.} \bibinfo{year}{2002}\natexlab{}.
\newblock \bibinfo{booktitle}{\emph{The news interview: Journalists and public
  figures on the air}}.
\newblock \bibinfo{publisher}{Cambridge University Press}.
\newblock


\bibitem[\protect\citeauthoryear{Danescu-Niculescu-Mizil, Cheng, Kleinberg, and
  Lee}{Danescu-Niculescu-Mizil et~al\mbox{.}}{2012}]%
        {Danescu-Niculescu-Mizil+Cheng+Kleinberg+Lee:12}
\bibfield{author}{\bibinfo{person}{Cristian Danescu-Niculescu-Mizil},
  \bibinfo{person}{Justin Cheng}, \bibinfo{person}{Jon Kleinberg}, {and}
  \bibinfo{person}{Lillian Lee}.} \bibinfo{year}{2012}\natexlab{}.
\newblock \showarticletitle{You Had Me at Hello: How Phrasing Affects
  Memorability}. In \bibinfo{booktitle}{\emph{Proceedings of ACL}}.
\newblock


\bibitem[\protect\citeauthoryear{Diakopoulos and Shamma}{Diakopoulos and
  Shamma}{2010}]%
        {diakopoulos2010characterizing}
\bibfield{author}{\bibinfo{person}{Nicholas~A. Diakopoulos} {and}
  \bibinfo{person}{David~A. Shamma}.} \bibinfo{year}{2010}\natexlab{}.
\newblock \showarticletitle{Characterizing Debate Performance via Aggregated
  Twitter Sentiment}. In \bibinfo{booktitle}{\emph{Proceedings of CHI}}.
\newblock


\bibitem[\protect\citeauthoryear{Dinardo}{Dinardo}{2010}]%
        {Dinardo:Microeconometrics:2010}
\bibfield{author}{\bibinfo{person}{John Dinardo}.}
  \bibinfo{year}{2010}\natexlab{}.
\newblock \showarticletitle{Natural Experiments and Quasi-Natural Experiments}.
\newblock In \bibinfo{booktitle}{\emph{Microeconometrics}}.
  \bibinfo{publisher}{Palgrave Macmillan UK}, \bibinfo{pages}{139--153}.
\newblock


\bibitem[\protect\citeauthoryear{Entman}{Entman}{1993}]%
        {Entman:JournalOfCommunication:1993}
\bibfield{author}{\bibinfo{person}{Robert~M. Entman}.}
  \bibinfo{year}{1993}\natexlab{}.
\newblock \showarticletitle{Framing: Toward Clarification of a Fractured
  Paradigm}.
\newblock \bibinfo{journal}{\emph{Journal of Communication}}
  \bibinfo{volume}{43}, \bibinfo{number}{4} (\bibinfo{year}{1993}),
  \bibinfo{pages}{51--58}.
\newblock


\bibitem[\protect\citeauthoryear{Fridkin, Kenney, Gershon, and
  Serignese~Woodall}{Fridkin et~al\mbox{.}}{2008}]%
        {Fridkin01012008}
\bibfield{author}{\bibinfo{person}{Kim~L. Fridkin}, \bibinfo{person}{Patrick~J.
  Kenney}, \bibinfo{person}{Sarah~Allen Gershon}, {and} \bibinfo{person}{Gina
  Serignese~Woodall}.} \bibinfo{year}{2008}\natexlab{}.
\newblock \showarticletitle{Spinning Debates: The Impact of the News Media's
  Coverage of the Final 2004 Presidential Debate}.
\newblock \bibinfo{journal}{\emph{The International Journal of Press/Politics}}
  \bibinfo{volume}{13}, \bibinfo{number}{1} (\bibinfo{year}{2008}),
  \bibinfo{pages}{29--51}.
\newblock


\bibitem[\protect\citeauthoryear{Geer}{Geer}{2012}]%
        {geer2012news}
\bibfield{author}{\bibinfo{person}{John~G. Geer}.}
  \bibinfo{year}{2012}\natexlab{}.
\newblock \showarticletitle{The News Media and the Rise of Negativity in
  Presidential Campaigns}.
\newblock \bibinfo{journal}{\emph{PS: Political Science \& Politics}}
  \bibinfo{volume}{45}, \bibinfo{number}{03} (\bibinfo{year}{2012}),
  \bibinfo{pages}{422--427}.
\newblock


\bibitem[\protect\citeauthoryear{Gidengil and Everitt}{Gidengil and
  Everitt}{2003}]%
        {gidengil2003talking}
\bibfield{author}{\bibinfo{person}{Elisabeth Gidengil} {and}
  \bibinfo{person}{Joanna Everitt}.} \bibinfo{year}{2003}\natexlab{}.
\newblock \showarticletitle{Talking Tough: Gender and Reported Speech in
  Campaign News Coverage}.
\newblock \bibinfo{journal}{\emph{Political communication}}
  \bibinfo{volume}{20}, \bibinfo{number}{3} (\bibinfo{year}{2003}),
  \bibinfo{pages}{209--232}.
\newblock


\bibitem[\protect\citeauthoryear{Golan}{Golan}{2007}]%
        {Golan:JournalismStudies:2007}
\bibfield{author}{\bibinfo{person}{Guy Golan}.}
  \bibinfo{year}{2007}\natexlab{}.
\newblock \showarticletitle{Inter-media Agenda Setting and Global News
  Coverage}.
\newblock \bibinfo{journal}{\emph{Journalism Studies}} \bibinfo{volume}{7},
  \bibinfo{number}{2} (\bibinfo{year}{2007}), \bibinfo{pages}{323--333}.
\newblock


\bibitem[\protect\citeauthoryear{Grand}{Grand}{2015}]%
        {grand2015emergence}
\bibfield{author}{\bibinfo{person}{Noah Grand}.}
  \bibinfo{year}{2015}\natexlab{}.
\newblock \emph{\bibinfo{title}{The Emergence of Newsworthiness: Inclusion,
  Exclusion and Inequality in Political News and Online Media}}.
\newblock \bibinfo{thesistype}{Ph.D. Dissertation}.
\newblock


\bibitem[\protect\citeauthoryear{Groseclose and Milyo}{Groseclose and
  Milyo}{2005}]%
        {groseclose2005measure}
\bibfield{author}{\bibinfo{person}{Tim Groseclose} {and}
  \bibinfo{person}{Jeffrey Milyo}.} \bibinfo{year}{2005}\natexlab{}.
\newblock \showarticletitle{A Measure of Media Bias}.
\newblock \bibinfo{journal}{\emph{The Quarterly Journal of Economics}}
  (\bibinfo{year}{2005}), \bibinfo{pages}{1191--1237}.
\newblock


\bibitem[\protect\citeauthoryear{Gross, Porter, and Wood}{Gross
  et~al\mbox{.}}{2017}]%
        {gross+etal:17}
\bibfield{author}{\bibinfo{person}{Kimberly Gross}, \bibinfo{person}{Ethan
  Porter}, {and} \bibinfo{person}{Thomas Wood}.}
  \bibinfo{year}{2017}\natexlab{}.
\newblock \bibinfo{title}{Presidential Debates in the Age of Partisan Media: A
  Field Experiment}.  (\bibinfo{year}{2017}).
\newblock
\newblock
\shownote{Available at SSRN.}


\bibitem[\protect\citeauthoryear{Hallin}{Hallin}{1992}]%
        {hallin1992sound}
\bibfield{author}{\bibinfo{person}{Daniel~C. Hallin}.}
  \bibinfo{year}{1992}\natexlab{}.
\newblock \showarticletitle{Sound Bite News: Television Coverage of Elections,
  1968--1988}.
\newblock \bibinfo{journal}{\emph{Journal of communication}}
  \bibinfo{volume}{42}, \bibinfo{number}{2} (\bibinfo{year}{1992}),
  \bibinfo{pages}{5--24}.
\newblock


\bibitem[\protect\citeauthoryear{Hillygus and Jackman}{Hillygus and
  Jackman}{2003}]%
        {Hillygus:AmericanJournalOfPoliticalScience:2003}
\bibfield{author}{\bibinfo{person}{D.~Sunshine Hillygus} {and}
  \bibinfo{person}{Simon Jackman}.} \bibinfo{year}{2003}\natexlab{}.
\newblock \showarticletitle{Voter Decision Making in Election 2000: Campaign
  Effects, Partisan Activation, and the {Clinton} Legacy}.
\newblock \bibinfo{journal}{\emph{American Journal of Political Science}}
  \bibinfo{volume}{47}, \bibinfo{number}{4} (\bibinfo{year}{2003}),
  \bibinfo{pages}{583--596}.
\newblock


\bibitem[\protect\citeauthoryear{Hwang, Gotlieb, Nah, and McLeod}{Hwang
  et~al\mbox{.}}{2007}]%
        {hwang2007applying}
\bibfield{author}{\bibinfo{person}{Hyunseo Hwang}, \bibinfo{person}{Melissa~R.
  Gotlieb}, \bibinfo{person}{Seungahn Nah}, {and} \bibinfo{person}{Douglas~M.
  McLeod}.} \bibinfo{year}{2007}\natexlab{}.
\newblock \showarticletitle{Applying a Cognitive-processing Model to
  Presidential Debate Effects: Postdebate News Analysis and Primed Reflection}.
\newblock \bibinfo{journal}{\emph{Journal of Communication}}
  \bibinfo{volume}{57}, \bibinfo{number}{1} (\bibinfo{year}{2007}),
  \bibinfo{pages}{40--59}.
\newblock


\bibitem[\protect\citeauthoryear{Iyengar and Hahn}{Iyengar and Hahn}{2009}]%
        {iyengar2009red}
\bibfield{author}{\bibinfo{person}{Shanto Iyengar} {and}
  \bibinfo{person}{Kyu~S. Hahn}.} \bibinfo{year}{2009}\natexlab{}.
\newblock \showarticletitle{Red media, Blue media: Evidence of Ideological
  Selectivity in Media Use}.
\newblock \bibinfo{journal}{\emph{Journal of Communication}}
  \bibinfo{volume}{59}, \bibinfo{number}{1} (\bibinfo{year}{2009}),
  \bibinfo{pages}{19--39}.
\newblock


\bibitem[\protect\citeauthoryear{Joachims}{Joachims}{2002}]%
        {Joachims:2002:OSE:775047.775067}
\bibfield{author}{\bibinfo{person}{Thorsten Joachims}.}
  \bibinfo{year}{2002}\natexlab{}.
\newblock \showarticletitle{Optimizing Search Engines Using Clickthrough Data}.
  In \bibinfo{booktitle}{\emph{Proceedings of KDD}}.
\newblock


\bibitem[\protect\citeauthoryear{Kim and Barnett}{Kim and Barnett}{1996}]%
        {kim+barnett:1996}
\bibfield{author}{\bibinfo{person}{Kyungmo Kim} {and}
  \bibinfo{person}{George~A. Barnett}.} \bibinfo{year}{1996}\natexlab{}.
\newblock \showarticletitle{The Determinants of International News Flow}.
\newblock \bibinfo{journal}{\emph{Communication Research}}
  \bibinfo{volume}{23}, \bibinfo{number}{3} (\bibinfo{year}{1996}),
  \bibinfo{pages}{323--352}.
\newblock


\bibitem[\protect\citeauthoryear{Lakoff}{Lakoff}{1975}]%
        {lakoff1975hedges}
\bibfield{author}{\bibinfo{person}{George Lakoff}.}
  \bibinfo{year}{1975}\natexlab{}.
\newblock \showarticletitle{Hedges: A Study in Meaning Criteria and the Logic
  of Fuzzy Concepts}.
\newblock \bibinfo{journal}{\emph{Journal of Philosophical Logic}}
  \bibinfo{volume}{2}, \bibinfo{number}{4} (\bibinfo{year}{1975}),
  \bibinfo{pages}{458--508}.
\newblock


\bibitem[\protect\citeauthoryear{Leskovec, Backstrom, and Kleinberg}{Leskovec
  et~al\mbox{.}}{2009}]%
        {leskovec2009meme}
\bibfield{author}{\bibinfo{person}{Jure Leskovec}, \bibinfo{person}{Lars
  Backstrom}, {and} \bibinfo{person}{Jon Kleinberg}.}
  \bibinfo{year}{2009}\natexlab{}.
\newblock \showarticletitle{Meme-tracking and the Dynamics of the News Cycle}.
  In \bibinfo{booktitle}{\emph{Proceedings of KDD}}.
\newblock


\bibitem[\protect\citeauthoryear{Lin, Bagrow, and Lazer}{Lin
  et~al\mbox{.}}{2011}]%
        {lin2011more}
\bibfield{author}{\bibinfo{person}{Yu-Ru Lin}, \bibinfo{person}{James~P.
  Bagrow}, {and} \bibinfo{person}{David Lazer}.}
  \bibinfo{year}{2011}\natexlab{}.
\newblock \showarticletitle{More Voices Than Ever? {Quantifying} Media Bias in
  Networks}. In \bibinfo{booktitle}{\emph{Proceedings of ICWSM}}.
\newblock


\bibitem[\protect\citeauthoryear{McCombs and Shaw}{McCombs and Shaw}{1972}]%
        {mccombs1972agenda}
\bibfield{author}{\bibinfo{person}{Maxwell~E. McCombs} {and}
  \bibinfo{person}{Donald~L. Shaw}.} \bibinfo{year}{1972}\natexlab{}.
\newblock \showarticletitle{The Agenda-setting Function of Mass Media}.
\newblock \bibinfo{journal}{\emph{Public Opinion Quarterly}}
  \bibinfo{volume}{36}, \bibinfo{number}{2} (\bibinfo{year}{1972}),
  \bibinfo{pages}{176--187}.
\newblock


\bibitem[\protect\citeauthoryear{McKinney and Carlin}{McKinney and
  Carlin}{2004}]%
        {mckinney2004political}
\bibfield{author}{\bibinfo{person}{Mitchell~S. McKinney} {and}
  \bibinfo{person}{Diana~B. Carlin}.} \bibinfo{year}{2004}\natexlab{}.
\newblock \showarticletitle{Political Campaign Debates}.
\newblock In \bibinfo{booktitle}{\emph{Handbook of political communication
  research}}. \bibinfo{pages}{203--234}.
\newblock


\bibitem[\protect\citeauthoryear{Nguyen, Boyd-Graber, and Resnik}{Nguyen
  et~al\mbox{.}}{2012}]%
        {Nguyen:2012:SHN:2390524.2390536}
\bibfield{author}{\bibinfo{person}{Viet-An Nguyen}, \bibinfo{person}{Jordan
  Boyd-Graber}, {and} \bibinfo{person}{Philip Resnik}.}
  \bibinfo{year}{2012}\natexlab{}.
\newblock \showarticletitle{{SITS}: A Hierarchical Nonparametric Model Using
  Speaker Identity for Topic Segmentation in Multiparty Conversations}. In
  \bibinfo{booktitle}{\emph{Proceedings of ACL}}.
\newblock


\bibitem[\protect\citeauthoryear{Nguyen, Boyd-Graber, Resnik, Cai, Midberry,
  and Wang}{Nguyen et~al\mbox{.}}{2014}]%
        {Nguyen:2014:MTC:2629710.2629739}
\bibfield{author}{\bibinfo{person}{Viet-An Nguyen}, \bibinfo{person}{Jordan
  Boyd-Graber}, \bibinfo{person}{Philip Resnik}, \bibinfo{person}{Deborah~A.
  Cai}, \bibinfo{person}{Jennifer~E. Midberry}, {and} \bibinfo{person}{Yuanxin
  Wang}.} \bibinfo{year}{2014}\natexlab{}.
\newblock \showarticletitle{Modeling Topic Control to Detect Influence in
  Conversations Using Nonparametric Topic Models}.
\newblock \bibinfo{journal}{\emph{Machine Learning}} \bibinfo{volume}{95},
  \bibinfo{number}{3} (\bibinfo{year}{2014}), \bibinfo{pages}{381--421}.
\newblock


\bibitem[\protect\citeauthoryear{Niculae, Suen, Zhang, Danescu-Niculescu-Mizil,
  and Leskovec}{Niculae et~al\mbox{.}}{2015}]%
        {Niculae:2015:QSP:2736277.2741688}
\bibfield{author}{\bibinfo{person}{Vlad Niculae}, \bibinfo{person}{Caroline
  Suen}, \bibinfo{person}{Justine Zhang}, \bibinfo{person}{Cristian
  Danescu-Niculescu-Mizil}, {and} \bibinfo{person}{Jure Leskovec}.}
  \bibinfo{year}{2015}\natexlab{}.
\newblock \showarticletitle{{QUOTUS}: The Structure of Political Media Coverage
  as Revealed by Quoting Patterns}. In \bibinfo{booktitle}{\emph{Proceedings of
  WWW}}.
\newblock


\bibitem[\protect\citeauthoryear{Patterson}{Patterson}{2016a}]%
        {patterson2016press}
\bibfield{author}{\bibinfo{person}{Thomas~E. Patterson}.}
  \bibinfo{year}{2016}\natexlab{a}.
\newblock \bibinfo{title}{News Coverage of the 2016 General Election: How the
  Press Failed the Voters}.
\newblock   (\bibinfo{year}{2016}).
\newblock


\bibitem[\protect\citeauthoryear{Patterson}{Patterson}{2016b}]%
        {patterson2016news}
\bibfield{author}{\bibinfo{person}{Thomas~E. Patterson}.}
  \bibinfo{year}{2016}\natexlab{b}.
\newblock \bibinfo{title}{News Coverage of the 2016 Presidential Primaries:
  Horce Race Reporting Has Consequences}.
\newblock   (\bibinfo{year}{2016}).
\newblock


\bibitem[\protect\citeauthoryear{Pennebaker, Francis, and Booth}{Pennebaker
  et~al\mbox{.}}{2007}]%
        {pennebaker2001linguistic}
\bibfield{author}{\bibinfo{person}{James~W. Pennebaker},
  \bibinfo{person}{Martha~E. Francis}, {and} \bibinfo{person}{Roger~J. Booth}.}
  \bibinfo{year}{2007}\natexlab{}.
\newblock \bibinfo{booktitle}{\emph{Linguistic Inquiry and Word Count: {LIWC}
  2007}}.
\newblock \bibinfo{type}{{T}echnical {R}eport}.
\newblock


\bibitem[\protect\citeauthoryear{Prabhakaran, Arora, and Rambow}{Prabhakaran
  et~al\mbox{.}}{2014}]%
        {prabhakaran2014staying}
\bibfield{author}{\bibinfo{person}{Vinodkumar Prabhakaran},
  \bibinfo{person}{Ashima Arora}, {and} \bibinfo{person}{Owen Rambow}.}
  \bibinfo{year}{2014}\natexlab{}.
\newblock \showarticletitle{Staying on Topic: An Indicator of Power in
  Political Debates}. In \bibinfo{booktitle}{\emph{Proceedings of EMNLP}}.
\newblock


\bibitem[\protect\citeauthoryear{Prat and Str\"{o}mberg}{Prat and
  Str\"{o}mberg}{2013}]%
        {prat2011political}
\bibfield{author}{\bibinfo{person}{Andrea Prat} {and} \bibinfo{person}{David
  Str\"{o}mberg}.} \bibinfo{year}{2013}\natexlab{}.
\newblock \showarticletitle{The Political Economy of Mass Media}.
\newblock In \bibinfo{booktitle}{\emph{Advances in Economics and Econometrics:
  Tenth World Congress}}. \bibinfo{publisher}{Cambridge University Press}.
\newblock


\bibitem[\protect\citeauthoryear{Rousseeuw}{Rousseeuw}{1987}]%
        {rousseeuw1987silhouettes}
\bibfield{author}{\bibinfo{person}{Peter~J. Rousseeuw}.}
  \bibinfo{year}{1987}\natexlab{}.
\newblock \showarticletitle{Silhouettes: A Graphical Aid to the Interpretation
  and Validation of Cluster Analysis}.
\newblock \bibinfo{journal}{\emph{{Journal of Computational and Applied
  Mathematics}}}  \bibinfo{volume}{20} (\bibinfo{year}{1987}),
  \bibinfo{pages}{53--65}.
\newblock


\bibitem[\protect\citeauthoryear{Rozin and Royzman}{Rozin and Royzman}{2001}]%
        {rozin2001negativity}
\bibfield{author}{\bibinfo{person}{Paul Rozin} {and} \bibinfo{person}{Edward~B.
  Royzman}.} \bibinfo{year}{2001}\natexlab{}.
\newblock \showarticletitle{Negativity Bias, Negativity Dominance, and
  Contagion}.
\newblock \bibinfo{journal}{\emph{Personality and Social Psychology Review}}
  \bibinfo{volume}{5}, \bibinfo{number}{4} (\bibinfo{year}{2001}),
  \bibinfo{pages}{296--320}.
\newblock


\bibitem[\protect\citeauthoryear{Shahaf, Horvitz, and Mankoff}{Shahaf
  et~al\mbox{.}}{2015}]%
        {Shahaf:2015:IJI:2783258.2783388}
\bibfield{author}{\bibinfo{person}{Dafna Shahaf}, \bibinfo{person}{Eric
  Horvitz}, {and} \bibinfo{person}{Robert Mankoff}.}
  \bibinfo{year}{2015}\natexlab{}.
\newblock \showarticletitle{Inside Jokes: Identifying Humorous Cartoon
  Captions}. In \bibinfo{booktitle}{\emph{Proceedings of KDD}}.
\newblock


\bibitem[\protect\citeauthoryear{Sim, Acree, Gross, and Smith}{Sim
  et~al\mbox{.}}{2013}]%
        {sim:2013:emnlp}
\bibfield{author}{\bibinfo{person}{Yanchuan Sim}, \bibinfo{person}{Brice~D.L.
  Acree}, \bibinfo{person}{Justin~H. Gross}, {and} \bibinfo{person}{Noah~A.
  Smith}.} \bibinfo{year}{2013}\natexlab{}.
\newblock \showarticletitle{Measuring Ideological Proportions in Political
  Speeches}. In \bibinfo{booktitle}{\emph{Proceedings of EMNLP}}.
\newblock


\bibitem[\protect\citeauthoryear{Simmons, Adamic, and Adar}{Simmons
  et~al\mbox{.}}{2011}]%
        {simmons2011memes}
\bibfield{author}{\bibinfo{person}{Matthew~P. Simmons},
  \bibinfo{person}{Lada~A. Adamic}, {and} \bibinfo{person}{Eytan Adar}.}
  \bibinfo{year}{2011}\natexlab{}.
\newblock \showarticletitle{Memes Online: Extracted, Subtracted, Injected, and
  Recollected}. In \bibinfo{booktitle}{\emph{Proceedings of ICWSM}}.
\newblock


\bibitem[\protect\citeauthoryear{Song, Liu, Fu, Liu, Wang, and Liu}{Song
  et~al\mbox{.}}{2016}]%
        {songlearning}
\bibfield{author}{\bibinfo{person}{Wei Song}, \bibinfo{person}{Tong Liu},
  \bibinfo{person}{Ruiji Fu}, \bibinfo{person}{Lizhen Liu},
  \bibinfo{person}{Hanshi Wang}, {and} \bibinfo{person}{Ting Liu}.}
  \bibinfo{year}{2016}\natexlab{}.
\newblock \showarticletitle{Learning to Identify Sentence Parallelism in
  Student Essays}. In \bibinfo{booktitle}{\emph{Proceedings of COLING}}.
\newblock


\bibitem[\protect\citeauthoryear{Stoer and Wagner}{Stoer and Wagner}{1997}]%
        {stoer1997simple}
\bibfield{author}{\bibinfo{person}{Mechthild Stoer} {and}
  \bibinfo{person}{Frank Wagner}.} \bibinfo{year}{1997}\natexlab{}.
\newblock \showarticletitle{A Simple Min-cut Algorithm}.
\newblock \bibinfo{journal}{\emph{{Journal of the ACM}}} \bibinfo{volume}{44},
  \bibinfo{number}{4} (\bibinfo{year}{1997}), \bibinfo{pages}{585--591}.
\newblock


\bibitem[\protect\citeauthoryear{Stolcke, Zheng, Wang, and Abrash}{Stolcke
  et~al\mbox{.}}{2011}]%
        {andreas2011srilm}
\bibfield{author}{\bibinfo{person}{Andreas Stolcke}, \bibinfo{person}{Jing
  Zheng}, \bibinfo{person}{Wen Wang}, {and} \bibinfo{person}{Victor Abrash}.}
  \bibinfo{year}{2011}\natexlab{}.
\newblock \showarticletitle{SRILM at Sixteen: Update and Outlook}. In
  \bibinfo{booktitle}{\emph{Proceedings of IEEE Automatic Speech Recognition
  and Understanding Workshop}}.
\newblock


\bibitem[\protect\citeauthoryear{Stroud}{Stroud}{2010}]%
        {Stroud:JournalOfCommunication:2010}
\bibfield{author}{\bibinfo{person}{Natalie~J. Stroud}.}
  \bibinfo{year}{2010}\natexlab{}.
\newblock \showarticletitle{Polarization and Partisan Selective Exposure}.
\newblock \bibinfo{journal}{\emph{Journal of Communication}}
  \bibinfo{volume}{60}, \bibinfo{number}{3} (\bibinfo{year}{2010}),
  \bibinfo{pages}{556--576}.
\newblock


\bibitem[\protect\citeauthoryear{Tan, Friggeri, and Adamic}{Tan
  et~al\mbox{.}}{2016a}]%
        {tan2016lost}
\bibfield{author}{\bibinfo{person}{Chenhao Tan}, \bibinfo{person}{Adrien
  Friggeri}, {and} \bibinfo{person}{Lada~A. Adamic}.}
  \bibinfo{year}{2016}\natexlab{a}.
\newblock \showarticletitle{Lost in Propagation? {Unfolding} News Cycles from
  the Source}. In \bibinfo{booktitle}{\emph{Proceedings of ICWSM}}.
\newblock


\bibitem[\protect\citeauthoryear{Tan and Lee}{Tan and Lee}{2016}]%
        {tan+lee:16tad}
\bibfield{author}{\bibinfo{person}{Chenhao Tan} {and} \bibinfo{person}{Lillian
  Lee}.} \bibinfo{year}{2016}\natexlab{}.
\newblock \bibinfo{title}{Talk it up or Play it down? {(Un)expected}
  Correlations between (De-)emphasis and Recurrence of Discussion Points in
  Consequential {U.S.} Economic Policy Meetings}.  (\bibinfo{year}{2016}).
\newblock
\newblock
\shownote{Presented in Text as Data.}


\bibitem[\protect\citeauthoryear{Tan, Lee, and Pang}{Tan et~al\mbox{.}}{2014}]%
        {Tan:ProceedingsOfAcl:2014}
\bibfield{author}{\bibinfo{person}{Chenhao Tan}, \bibinfo{person}{Lillian Lee},
  {and} \bibinfo{person}{Bo Pang}.} \bibinfo{year}{2014}\natexlab{}.
\newblock \showarticletitle{The Effect of Wording on Message Propagation:
  Topic- and Author-controlled Natural Experiments on Twitter}. In
  \bibinfo{booktitle}{\emph{Proceedings of ACL}}.
\newblock


\bibitem[\protect\citeauthoryear{Tan, Niculae, Danescu-Niculescu-Mizil, and
  Lee}{Tan et~al\mbox{.}}{2016b}]%
        {Tan:2016:WAI:2872427.2883081}
\bibfield{author}{\bibinfo{person}{Chenhao Tan}, \bibinfo{person}{Vlad
  Niculae}, \bibinfo{person}{Cristian Danescu-Niculescu-Mizil}, {and}
  \bibinfo{person}{Lillian Lee}.} \bibinfo{year}{2016}\natexlab{b}.
\newblock \showarticletitle{Winning Arguments: Interaction Dynamics and
  Persuasion Strategies in Good-faith Online Discussions}. In
  \bibinfo{booktitle}{\emph{Proceedings of WWW}}.
\newblock


\bibitem[\protect\citeauthoryear{Tsfati}{Tsfati}{2003}]%
        {Tsfati01072003}
\bibfield{author}{\bibinfo{person}{Yariv Tsfati}.}
  \bibinfo{year}{2003}\natexlab{}.
\newblock \showarticletitle{Debating the Debate: The Impact of Exposure to
  Debate News Coverage and Its Interaction with Exposure to the Actual Debate}.
\newblock \bibinfo{journal}{\emph{The International Journal of Press/Politics}}
  \bibinfo{volume}{8}, \bibinfo{number}{3} (\bibinfo{year}{2003}),
  \bibinfo{pages}{70--86}.
\newblock


\bibitem[\protect\citeauthoryear{Wells, Shah, Pevehouse, Yang, Pelled, Boehm,
  Lukito, Ghosh, and Schmidt}{Wells et~al\mbox{.}}{2016}]%
        {Wells:PoliticalCommunication:2016}
\bibfield{author}{\bibinfo{person}{Chris Wells}, \bibinfo{person}{Dhavan~V.
  Shah}, \bibinfo{person}{Jon~C. Pevehouse}, \bibinfo{person}{JungHwan Yang},
  \bibinfo{person}{Ayellet Pelled}, \bibinfo{person}{Frederick Boehm},
  \bibinfo{person}{Josephine Lukito}, \bibinfo{person}{Shreenita Ghosh}, {and}
  \bibinfo{person}{Jessica~L Schmidt}.} \bibinfo{year}{2016}\natexlab{}.
\newblock \showarticletitle{{How Trump Drove Coverage to the Nomination: Hybrid
  Media Campaigning}}.
\newblock \bibinfo{journal}{\emph{Political Communication}}
  (\bibinfo{year}{2016}).
\newblock


\bibitem[\protect\citeauthoryear{Zhang, Kumar, Ravi, and
  Danescu-Niculescu-Mizil}{Zhang et~al\mbox{.}}{2016}]%
        {Zhang:ProceedingsOfNaacl:2016}
\bibfield{author}{\bibinfo{person}{Justine Zhang}, \bibinfo{person}{Ravi
  Kumar}, \bibinfo{person}{Sujith Ravi}, {and} \bibinfo{person}{Cristian
  Danescu-Niculescu-Mizil}.} \bibinfo{year}{2016}\natexlab{}.
\newblock \showarticletitle{Conversational Flow in {Oxford-style} Debates}. In
  \bibinfo{booktitle}{\emph{Proceedings of NAACL (short papers)}}.
\newblock


\end{thebibliography}

\end{document}